\documentclass[aps,twocolumn,showpacs,prb,reprint]{revtex4-1}

\usepackage{amsmath}
\usepackage{amsfonts}
\usepackage{amssymb}
\usepackage{graphicx}
\usepackage{color}
\usepackage{xcolor}
\usepackage[titletoc,toc]{appendix}

\usepackage{tikz}
\usetikzlibrary{positioning,arrows}
\usetikzlibrary{decorations.pathmorphing}
\usetikzlibrary{decorations.markings}
\usetikzlibrary{calc}
\usetikzlibrary{patterns}
\usetikzlibrary{shapes.misc}
\tikzset{
phys/.style={thick, postaction={decorate}, decoration={markings, mark=at position .7 with {\arrow[]{triangle 45}}}},
res/.style={thick, decorate, decoration={snake,segment length=7, amplitude=1.5}},
arrow/.style={thick, draw=white, postaction={decorate}, decoration={markings, mark=at position .6 with {\arrow[black]{triangle 45}}}}
 }
 
\newcommand{\dd}{\mathrm{d}}
\newcommand{\ee}{\mathrm{e}}

\newcommand{\rr}{\mathbf{r}}

\newcommand{\kk}{\mathbf{k}}
\newcommand{\qq}{\mathbf{q}}
\newcommand{\pp}{\mathbf{p}}
\newcommand{\meff}{\widetilde{m}}
\newcommand{\geff}{\widetilde{g}}
\newcommand{\leff}{\widetilde{\lambda}}
\newcommand{\res}{\widetilde{\phi}}

\begin{document}

\title{Universal short-time dynamics: boundary functional renormalization group for a temperature quench}
\author{Alessio Chiocchetta}
\affiliation{SISSA --- International School for Advanced Studies and INFN, via Bonomea 265, 34136 Trieste, Italy}
\author{Andrea Gambassi}
\affiliation{SISSA --- International School for Advanced Studies and INFN, via Bonomea 265, 34136 Trieste, Italy}
\author{Sebastian Diehl}
\affiliation{Institut f\"{u}r Theoretische Physik, Universit\"{a}t zu K\"{o}ln, D-50937 Cologne, Germany}
\affiliation{Institut f\"{u}r Theoretische Physik, TU Dresden, D-01062 Dresden, Germany}
\author{Jamir Marino}
\affiliation{Institut f\"{u}r Theoretische Physik, Universit\"{a}t zu K\"{o}ln, D-50937 Cologne, Germany}
\affiliation{Institut f\"{u}r Theoretische Physik, TU Dresden, D-01062 Dresden, Germany}

\begin{abstract}
We present a method to calculate short-time non-equilibrium universal exponents within the functional renormalization-group scheme. As an example, we consider the classical critical dynamics of the relaxational model A after a quench of the temperature of the system and calculate the initial-slip exponent which characterizes the non-equilibrium universal short-time behaviour of both the order parameter and correlation functions. The value of this exponent is found to be consistent with the result of a perturbative dimensional expansion and of Monte Carlo simulations in three spatial dimensions. 
\end{abstract}


\date{\today}
\maketitle

\section{Introduction}
\label{sec:introduction}

The quest for non-equilibrium collective properties in macroscopic classical and quantum systems lies at the forefront of modern statistical physics. In fact, when macroscopic systems are driven out of equilibrium, they exhibit a variety of novel and fascinating phenomena which have no counterpart in equilibrium: when these collective properties are insensitive to the microscopic details of the system, \emph{non-equilibrium universality} emerges. Instances of it are found in both classical and quantum systems. For the former, examples are provided by relaxational models~\cite{Janssen1989,Janssen1992,Calabrese2005,Henkelbook2,Tauberbook2014}, driven-diffusive~\cite{Zia1995} and reaction-diffusion systems (both in stationary and transient regime)~\cite{Doi1976,Peliti1985,Cardy1996,Baumann2007}, directed percolation~\cite{Obukhov1980,Hinrichsen2000,Henkelbook1}, self-organized criticality~\cite{Jensen1998}, and roughening phenomena~\cite{Kardar1986,Corwin2012}. 
In quantum many-body systems, non-equilibrium universality was predicted in open electronic systems~\cite{Mitra2006}, noise-driven phase transitions~\cite{DallaTorreDemler2012}, superfluid turbulence and non-thermal fixed points of Bose gases~\cite{Berges2007, Berges2008, Berges2009, Scheppach2010,Nowak2011,Nowak2012,Schole2012,Berges2012,Orioli2015,Langen2016}, 
driven-dissipative quantum-optical platforms~\cite{DallaTorreDiehl2013,Sieberer2013,Altman2015,Marino2016,Marino2016b}, aging dynamics of isolated~\cite{Chiocchetta2015,Maraga2015,Chiocchetta2016} and open~\cite{Gagel2014,Gagel2015,Buchhold2014} quantum systems, dynamical phase transitions~\cite{Eckstein2009,Schiro2010,Sciolla2010,Schiro2011,Sciolla2011,Gambassi2011,Sciolla2013,Chandran2013,Smacchia2015,Maraga2016}, and in the statistics of the work done upon quenching~\cite{Gambassi2012,Sotiriadis2013}.

The theoretical investigation of non-equilibrium universality is, however, considerably more challenging than its equilibrium counterpart, since one cannot rely in general on the minimization of thermodynamic potentials or exploit fluctuation-dissipation relations~\cite{Kubo1966,Janssen1979,Janssen1992,Foini2011,Foini2012,Sieberer2015b}, which constrain static and dynamical properties in equilibrium systems. Accordingly, a systematic description of  non-equilibrium universality calls for the introduction of novel theoretical tools.  
 
The response-function formalism (also known as MSRJD formalism)~\cite{Martin1973,Janssen1976,DeDominicis1976,DeDominicis1978,Janssen1979,Janssen1992} provides a practical framework for  a systematic classification of equilibrium critical dynamics~\cite{Hohenberg1977,Tauberbook2014} based on a renormalization-group (RG) approach, which was also succesfully used  to study non-equilibrium classical critical systems~\cite{Janssen1989,Calabrese2005,Tauberbook2014}. A similar formalism, based on the Schwinger-Keldysh functional~\cite{Berges2004b,Kamenevbook2011,Altlandbook2010,Berges2015}, is correspondingly used for investigating  non-equilibrium criticality in quantum systems.  
While the typical RG scheme used for studying  non-equilibrium universality is based on the dimensional expansion~\cite{Tauberbook2014}, the functional renormalization group (FRG) has been recently introduced for the investigation of non-equilibrium classical~\cite{Canet2004,Canet2005,Canet2010} and quantum systems~\cite{Sieberer2013,Sieberer2014,Mathey2015,Sieberer2015,Marino2016}, where it turned out to be effective in providing quantitative predictions which are out of reach of low-order dimensional expansions. 
FRG methods have been used, so far,  to investigate both the universal critical properties of non-equilibrium \emph{stationary} states of classical and quantum statistical systems, and the non-equilibrium real-time evolution of small quantum systems coupled to an environment~\cite{Gezzi2007,Jakobs2007,Pletyukhov2010,Kessler2010,Kennes2012,Metzner2012, Kennes2013,Kennes2013b}, with few notable exceptions concerning the non-equilibrium dynamics of many-body systems~\cite{Gasenzer2008, Kloss2011, Hick2012}.
 
In this paper, we introduce a  FRG scheme to address the non-equilibrium dynamics of classical  systems quenched close to a critical point: specifically, we consider the so-called stochastic model A~\cite{Hohenberg1977,Tauberbook2014} after the temperature of the thermal bath (which provides the thermal noise) has been quenched to the critical value. A concrete lattice realization of a system belonging to this universality class is the classical Ising model with non-conserved, i.e., spin-flip, dynamics. In fact, the non-equilibrium dynamics of this model exhibits a universal short-time behaviour~\cite{Janssen1989,Janssen1992,Calabrese2005}, which is revealed, e.g., in the scaling form of correlation functions and of the global magnetization, and is characterized by a new critical exponent, the so-called initial-slip exponent $\theta$. This universal quantity was first calculated at the second order in the dimensional $\epsilon$-expansion in Ref.~\onlinecite{Janssen1989}, and subsequently determined via numerical simulations (see Ref.~\onlinecite{Calabrese2005} for a summary). Here, we show how to calculate the exponent $\theta$ by implementing FRG within the response function formalism.

The presentation is organized as follows: in Sec.~\ref{sec:modelA} we introduce model A and the scaling form of correlation functions and of the order parameter after a critical quench. 
In Sec.~\ref{sec:RG-quench}, the FRG scheme is introduced for a quench, after rephrasing the Langevin dynamics of model A in a functional setting. In Sec.~\ref{sec:truncation-symmetric} we detail the results of our analysis for a simple ansatz of the effective action, benchmarking the method with the available predictions based on the analytical first-order (dimensional) $\epsilon$-expansion reported in the literature~\cite{Janssen1989}. 
In Sec.~\ref{sec:truncation-ordered}, we introduce an improved ansatz and we discuss and compare its results with those of a second-order (dimensional) $\epsilon$-expansion and of numerical Monte Carlo simulations. 
Finally, in Sec.~\ref{sec:concls} we provide an overview of potential applications of our approach to classical and quantum systems. 
All the relevant details of the calculations are reported in a number of Appendices.

\section{Critical quench of model A}  
\label{sec:modelA}
The so-called model A~\cite{Hohenberg1977,Tauberbook2014} captures the universal aspects of the relaxational dynamics of a classical system belonging to the Ising universality class and coupled to a thermal bath. This model prescribes an effective dynamics for the coarse-grained order parameter (i.e., the local magnetization), described by the classical field $\varphi \equiv \varphi(\rr,t)$ and evolving according to the Langevin equation
\begin{equation}
\label{eq:langevin}
\dot{\varphi} = - \Omega\frac{\delta \mathcal{H}}{\delta\varphi} + \zeta,
\end{equation}
where $\Omega$ is the diffusion coefficient, $\zeta$ is a zero-mean Markovian and Gaussian noise with correlation $\langle \zeta(\rr,t)\zeta(\rr',t') \rangle = 2 \Omega\, T \delta^{(d)}(\rr-\rr')\delta(t-t')$, describing the thermal fluctuations induced by the bath at temperature $T$ (measured in units of Boltzmann constant); $\mathcal{H}$ is given by
\begin{equation}
\label{eq:Hamiltonian}
\mathcal{H} = \int_\rr \left[ \frac{1}{2}(\nabla \varphi)^2 + \frac{\tau}{2}\varphi^2 + \frac{g}{4!}\varphi^4\right ],
\end{equation}
where $\int_\rr \equiv \int \dd^d r$ with $d$ the spatial dimensionality, $\tau$ parametrizes the distance from the critical point and $g\geq 0$ controls the strength of the interaction. The parameter $\tau$ depends on $T$ and it takes a critical value $\tau_c$ at the critical temperature $T=T_c$.  

We assume that the system is prepared at $t=t_0$ in the high-temperature phase with $T\to +\infty$ and an external magnetic field $h_0$, i.e., that the initial condition $\varphi(\rr,t=t_0) = \varphi_0(\rr)$ is a random field with probability distribution $P_0[\varphi_0]$ given by 
\begin{equation}
\label{eq:initial-probability}
P_0[\varphi_0] \propto \exp\left[-\int_\rr \frac{\tau_0}{2}(\varphi_0 -h_0)^2\right].
\end{equation}
Equation~\eqref{eq:initial-probability} implies that the initial field $\varphi_0$, with average  $\langle \varphi_0(\rr) \rangle = h_0(\rr)$, is characterized by short-range correlations
\begin{equation}
\langle \left[\varphi_0(\rr) -h_0(\rr)\right] \left[\varphi_0(\rr')-h_0(\rr')\right]\rangle=\tau_0^{-1}\delta^{(d)}(\rr-\rr'),
\end{equation}
where $1/\tau_0$ is the correlation length of the order parameter $\varphi_0(\mathbf{r})$ at $t=t_0$.
We recall that the correlation function $G_C$ is defined as~\cite{Tauberbook2014}
\begin{equation}
\label{eq:GK-definition}
G_C(\rr,t,t') = \langle \varphi(\rr,t)\varphi(\mathbf{0},t') \rangle,
\end{equation}
where $\langle\dots\rangle$ denotes the average over the dynamics generated by Eq.~\eqref{eq:langevin}, which includes averaging over both the initial condition $\varphi_0$ and the realizations of the noise $\zeta$. The response function $G_R$ is defined as the linear response to an external field $h(\rr,t)$, which couples linearly to $\varphi$ and which modifies 
the Hamiltonian $\mathcal{H}$ in Eq.~\eqref{eq:Hamiltonian} as $\mathcal{H}_h = \mathcal{H} - \int_\rr h\varphi$; specifically, we have
\begin{equation}
\label{eq:GR-definition}
G_R(\rr,t,t') \equiv \frac{\delta \langle \varphi(\rr,t)\rangle_h }{\delta h(\mathbf{0},t')} \biggr|_{h=0},
\end{equation}
where $\langle\dots\rangle_h$ denotes the average over the dynamics generated by Eq.~\eqref{eq:langevin} with the Hamiltonian $\mathcal{H}_h$. Note that in Eqs.~\eqref{eq:GK-definition} and~\eqref{eq:GR-definition} we made use of the spatial translational invariance of the dynamical equation~\eqref{eq:langevin}, as $G_C$ and $G_R$ only depend on the distance between the two spatial points involved in these equations. Accordingly, one can take the Fourier transform with respect to $\rr$ and express more conveniently $G_{C,R}$ in wave-vector space.

We assume that the temperature $T$ of the bath takes the critical value $T_c$ for $t>t_0$, so that the system will eventually relax to a critical equilibrium state. As a consequence of being at criticality, this relaxation dynamics exhibits self-similar properties, signalled by the emergence of a scaling behaviour referred to as aging: for example, correlation and response functions in momentum space read~\cite{Janssen1989,Calabrese2005,Gambassi2008}, after a quench occurring at $t=t_0$,
\begin{subequations}
\label{eq:greens-scaling}
\begin{align}
G_R(q, t, t' ) & \simeq q^{-2 + \eta + z} \left( \frac{t}{t'} \right)^\theta\,\mathcal{G}_R(q^zt), \label{eq:GR-scaling}\\
G_C(q, t, t') & \simeq q^{-2 + \eta} \left( \frac{t}{t'} \right)^{\theta-1}\,\mathcal{G}_C(q^zt), \label{eq:GK-scaling}
\end{align}
\end{subequations}
with $\eta$ the anomalous dimension~\cite{ZinnJustinbook,Goldenfeldbook}, $z$ the dynamical critical exponent~\cite{Mabook,Tauberbook2014}, and $\mathcal{G}_{R,C}(x)$ scaling functions. The scaling forms~\eqref{eq:greens-scaling} are valid for $h_0 =0$,  $t' \ll t$ and $t '\to t_m$, where $t_m$ is a microscopic time which depends on the specific details of the underlying microscopic model. The dynamics at times shorter than $t_m$ has a non-universal character and it depends on the material properties of the system. 
The scaling forms~\eqref{eq:greens-scaling} are characterized by the so-called initial-slip exponent $\theta$, which is generically independent of the static critical exponents $\eta, \nu$~\cite{ZinnJustinbook,Goldenfeldbook} and of the dynamical critical exponent $z$ characterizing the equilibrium dynamics of model A. The physical origin of $\theta$ can be eventually traced back to the (transient) violation of detailed balance due to the breaking of the time-translational invariance induced by the quench~\cite{Janssen1989}.   

\begin{figure}[h!]
\centering
\includegraphics[width=8.5cm]{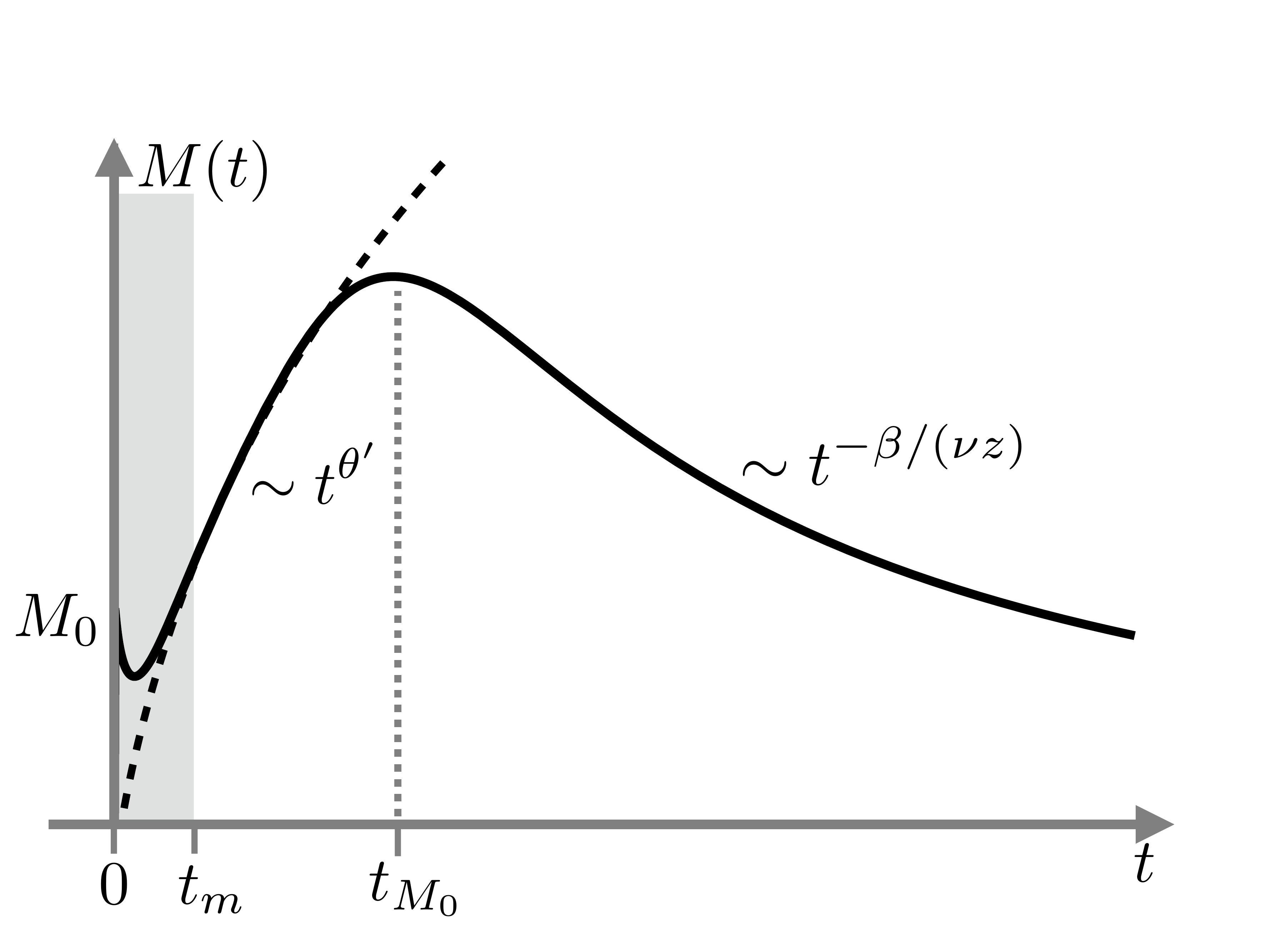}
 \caption{(Color online). Sketch of the time evolution of the magnetization $M(t)$ after a quench at $t=t_0$ from a disordered initial state with a small value $M_0$ of the magnetization to the critical temperature. The grey area indicates the time interval up to $t_m$ within which the dynamics does not display universal features.} 
\label{fig:fig1}
\end{figure}

In the presence of a non-vanishing initial homogeneous external field $h_0$, the evolution of the magnetization $M(t) \equiv \langle \varphi(\rr,t) \rangle$ displays an interesting non-equilibrium evolution. In fact, for $t \gg t_m$, it follows the scaling form~\cite{Janssen1989} 
\begin{equation}
\label{eq:M-scaling}
M(t) = M_0\, t^{\theta'} \mathcal{F}\left(M_0\, t^{\theta' + \beta/(\nu z)}\right),
\end{equation}
where $\theta' = \theta + (2-z-\eta)/z$, $\beta$ is the equilibrium critical exponent of the magnetization~\cite{ZinnJustinbook,Goldenfeldbook}, $M_0 \equiv h_0$ is the initial value of the magnetization and $\mathcal{F}(x)$ is a function with the following asymptotic properties:
\begin{equation}
\mathcal{F}(x) \approx 
\begin{cases}
x^{-1} & \text{for} \quad x\to \infty, \\
1 & \text{for} \quad x \to 0. \\ 
\end{cases}
\end{equation}
Accordingly, $M(t)$ exhibits the non-monotonic behavior depicted in Fig.~\ref{fig:fig1}: for times $t \lesssim t_{M_0} \propto M_0^{1/[\theta' + \beta/(\nu z)]}$  it grows as an algebraic function with the non-equilibrium exponent $\theta'$, while for $t \gtrsim t_{M_0}$ it relaxes towards its equilibrium value $M_\text{eq} = 0$, with an algebraic decay controlled by a combination of universal equilibrium (static and dynamic) critical exponents. 
%
%

\subsection{Gaussian approximation}
\label{sec:gaussian}
In the absence of interaction  ($g=0$), Eq.~\eqref{eq:langevin} is linear and therefore it is possible to calculate exactly the correlation and response functions. By solving Eq.~\eqref{eq:langevin} with $g=0$ and $h_0 = 0$, based on the definitions~\eqref{eq:GK-definition} and~\eqref{eq:GR-definition}, one finds, after a Fourier transform in space with wavevector  $q$,
\begin{align}
G_{0R}(q,t,t') & = \vartheta(t-t') \ee^{-\Omega\omega_q(t-t')}, \label{eq:GR-gaussian} \\
G_{0C}(q,t,t') & = \frac{T}{\omega_q} \Big[ \ee^{-\Omega\omega_q|t-t'|}+ \nonumber\\
&+\left(\frac{\omega_q}{T\tau_0} -1\right) \ee^{-\Omega\omega_q(t+t'-2t_0)}\Big], 
\label{eq:GK-gaussian} 
\end{align}
where $\omega_q=q^2+\tau$ is the dispersion relation, $\vartheta(t)$ is the Heaviside step function and $t_0$ is the time at which the quench occurs. The subscript $0$ in $G_{0C,0R}$ indicates that these expressions refer to the Gaussian approximation. Notice that, while $G_{0R}$ is a time-translational invariant function, as it depends only on the difference of times $t-t'$, $G_{0C}$ breaks time-translational invariance. However, by taking the initial time $t_0 \to -\infty$ and as long as $\omega_q \neq 0$, $G_{0C}$ recovers its equilibrium time-translational invariant form~\cite{Tauberbook2014}: this is, in fact, a consequence of the relaxational nature of model A, which erases at long times the information about the initial state.
In the presence of a non-vanishing initial homogeneous external field $h_0$, it is also possible to calculate exactly the evolution of the magnetization $M(t)$, i.e., 
\begin{equation}
\label{eq:M-gaussian}
M(t) = M_0\ee^{-\Omega \tau (t-t_0)},
\end{equation}
which vanishes exponentially fast in time for $\tau>0$, while it keeps its initial value $M_0 = h_0$ for $\tau=0$.

Within this Gaussian approximation, the dynamics~\eqref{eq:langevin} becomes critical for $\tau=0$: in this case, by comparing Eqs.~\eqref{eq:GR-gaussian}, \eqref{eq:GK-gaussian} and~\eqref{eq:M-gaussian} with Eqs.~\eqref{eq:GR-scaling}, \eqref{eq:GK-scaling} and Eq.~\eqref{eq:M-scaling}, one finds $\theta=0$, $\eta=0$  and $z=2$.

As a result of a having finite interaction strength $g \neq 0$, the Gaussian value of the initial-slip exponent acquires sizeable corrections~\cite{Janssen1989,Calabrese2005}. In Sec.~\ref{sec:RG-quench} we introduce a functional renormalization group formalism, which we employ in Secs.~\ref{sec:truncation-symmetric} and~\ref{sec:truncation-ordered} in order to calculate the resulting value of $\theta$. 

\section{Functional renormalization group for a quench}
\label{sec:RG-quench}
In general, breaking translational invariance in space and/or time prevents the use of ordinary computational strategies of FRG~\cite{Berges2002}, which are primarily based on writing the corresponding flow equations in Fourier space, where they acquire a particularly simple form; accordingly one has to resort to more advanced techniques\cite{Andergassen2004, Enss2005, Moore2015, Andergassen2006}. In this Section, we show how the case of broken time-translational invariance can be successfully and effectively studied.

\subsection{Response functional and FRG equation}
\label{sec:response-functional}
The Langevin formulation of model A in Eq.~\eqref{eq:langevin} can be converted into a functional form by using the response functional~\cite{Martin1973,Janssen1976,DeDominicis1976,DeDominicis1978,Janssen1979,Janssen1992,Tauberbook2014}. The corresponding action is given by
\begin{equation}
\label{eq:microscopic-action}
S[\varphi,\widetilde{\varphi}] = S_0[\varphi_0,\widetilde{\varphi}_0] + \int_\rr \int_{t_0}^{+\infty}\dd t\, \widetilde{\varphi} \left(\dot{\varphi} + \Omega\frac{\delta \mathcal{H}}{\delta \varphi} - \Omega T \widetilde{\varphi}\right),
\end{equation}
where $\widetilde{\varphi} = \widetilde{\varphi}(\rr,t)$ is the so-called response field, while $\widetilde{\varphi}_0 = \widetilde{\varphi}(\rr,t=t_0)$. The averages of quantities $O[\varphi,\widetilde{\varphi}]$ can thus be calculated via a functional integration as~\cite{Tauberbook2014}
\begin{equation}
\langle O[\varphi,\widetilde{\varphi}] \rangle=\int \mathcal{D}\varphi \mathcal{D}\widetilde{\varphi}~O[\varphi,\widetilde{\varphi}] e^{-S[\varphi,\widetilde{\varphi}]}.
\end{equation}
The action $S_0[\varphi_0,\widetilde{\varphi}_0]$ contains information about the initial state and can be derived by including the initial probability distribution~\eqref{eq:initial-probability} into the functional description~\cite{Janssen1989,Janssen1992,Calabrese2005}. We postpone the discussion of its precise form to Sec.~\ref{sec:QFRG}. 
The quench occurs at time $t_0$: if one is  interested only in the stationary properties of model A, the limit $t_0 \to -\infty$ can be taken, thus recovering a full time-translational invariant behaviour, as discussed in Sec.~\ref{sec:recovering-equilibrium}.

In order to implement the FRG~\cite{Berges2002, Delamotte2007}, it is necessary to supplement the action $S[\varphi,\widetilde{\varphi}]$ with a cutoff function $R_k(q)$, and to derive the one-particle irreducible effective action $\Gamma[\phi,\widetilde{\phi}]$  as the Legendre transform of the generating function associated to $S_k[\varphi,\widetilde{\varphi}]$ (see App.~\ref{app:wetterich}, in particular Eq.~\eqref{eq:Gamma-definition}). $R_k(q)$ is introduced as a quadratic term in the modified action $S_k[\Psi] \equiv S[\Psi] + \Delta S_k[\Psi]$, where $\Delta S_k[\Psi] =\int_{t,\rr} \Psi^\text{t} \sigma \Psi R_k/2$ with the Pauli matrix $\sigma =\left(\begin{smallmatrix} 0 & 1 \\1 & 0\end{smallmatrix}\right)$ acting in the two dimensional space of the variables $\varphi$ and $\widetilde{\varphi}$, encoded in  $\Psi^\text{t} = (\varphi,\widetilde{\varphi})$.
The cutoff function $R_k$ as a function of $k$ is characterized by the following limiting behaviours~\cite{Litim2000, Berges2002, Delamotte2007}:
\begin{equation}
R_k(q) \simeq
\begin{cases}
  \Lambda^2 & \text{for} \quad k\to\Lambda, \\
0 & \text{for} \quad k \to 0, \\ 
\end{cases}
\end{equation}
where $\Lambda$ is the ultraviolet cutoff of the model. Correspondingly, the effective action $\Gamma_k$ can be  interpreted as an action which interpolates between the microscopic one $S[\varphi,\widetilde{\varphi}]$ for $k \to \Lambda$,  and the long-distance effective one for $k\to 0$.
When the fluctuations of the order parameter are integrated out in order to evaluate the effective action, the effect of $R_k$ is to supplement slow modes with an effective $k$-dependent quadratic term (a mass, in field-theoretical language), allowing a smooth approach to the critical point, when the effective low-energy action is recovered for $k\to 0$. More specifically, for momenta $q\lesssim k$ the mass of the critical modes becomes proportional to $R_k(q)\simeq k^2$ (as we detail in Eq. \eqref{eq:modified-dispersion}), and this regularizes the infrared divergences of RG loop corrections, occurring at criticality when $q \to 0$ (see e.g. Refs. ~\onlinecite{Litim2000, Berges2002, Delamotte2007}). As a consequence of the introduction of the regulator $R_k$, the $k$-dependent effective action $\Gamma_k$ can then also be regarded as an action which has been coarse-grained on a spatial volume $ k^{-d}$. 

As discussed in App.~\ref{app:wetterich}, the flow equation for $\Gamma_k$ upon varying the coarse-graining scale $k$ is given by~\cite{Wetterich1993,Berges2002}
\begin{equation}
\label{eq:wetterich}
\frac{\dd \Gamma}{\dd k} = \frac{1}{2} \int_x  \text{tr}\left[ \vartheta(t-t_0)G(x,x)\frac{\dd R}{\dd k}\sigma\right],
\end{equation}
where, in order to simplify the notation, we no longer indicate explicitly the dependence on $k$ of $\Gamma$ and $R$, while we defined $x \equiv (\rr,t)$, $\int_x \equiv \int \dd^dr\int_{t_0}^{+\infty} \dd t$. The matrix $G(x,x')$ is defined as
\begin{equation}
\label{eq:G-definition}
G(x,x') = \left(\Gamma^{(2)} + R\,\sigma\right)^{-1}(x,x'), 
\end{equation}
where the inverse of the matrix on the r.h.s.~is taken with respect to spatial and temporal variables, as well as to the internal matrix structure. 
The kernel $\Gamma^{(2)}(x,x')$ is the second variation of the effective action $\Gamma$ with respect to the fields, i.e., 
\begin{equation}
\label{eq:Gamma-second-variation}
\Gamma^{(2)}(x,x')= 
\begin{pmatrix}
\displaystyle{\frac{\delta^2\Gamma}{\delta\phi(x)\delta\phi(x')}} & \displaystyle{\frac{\delta^2\Gamma}{\delta\res(x)\delta\phi(x')}} \\[4mm]
\displaystyle{\frac{\delta^2\Gamma}{\delta\phi(x)\delta\res(x')}} & \displaystyle{\frac{\delta^2\Gamma}{\delta\res(x)\delta\res(x')}}
\end{pmatrix}.
\end{equation}
While Eq.~\eqref{eq:wetterich} is exact, it is generally not possible to solve it. Accordingly, one has to resort to approximation schemes which render Eq.~\eqref{eq:wetterich} amenable to analytic and numerical calculations.
A first step in this direction is to provide an ansatz for the form of the effective action $\Gamma$ which, once inserted into Eq.~\eqref{eq:wetterich}, results in a set of coupled non-linear differential equations for the couplings which parametrize it. In fact, any coupling $g_{n,l}\phi^n\res^l/(n!\, l!)$ (with $n$ and $l$ positive integers) appearing in $\Gamma$ corresponds to a term of its vertex expansion~\cite{Amit/Martin-Mayor,Canet2007,Sieberer2014}, as
\begin{equation}
\label{eq:vertexp-maintext}
g_{n,l} = \frac{\delta^{l+n} \Gamma}{\delta\phi^n\delta\res^l}\biggr|_{\substack{\res =0 \\ \phi =\phi_m}}, 
\end{equation}
where the derivatives of $\Gamma$ are evaluated at some homogeneous field configurations $\res =0 $ and $\phi = \phi_m$. The field $\phi_m$, referred to as background field, is typically chosen as the minimum of the effective action $\Gamma$.  

In this work, we  consider the following ansatz for  model A:
\begin{align}
\label{eq:effective-action}
\Gamma[\phi,\res]	
& = \Gamma_0[\phi_0,\res_0] \nonumber \\
& \quad + \int_x \vartheta(t-t_0)\res \left( Z \dot{\phi} + K \nabla^2\phi + \frac{\partial \mathcal{U}}{\partial \phi} - D \res\right).
\end{align}
The boundary action $\Gamma_0[\phi_0,\res_0]$ accounts for the initial conditions and its form will be discussed in detail in Sec.~\ref{sec:QFRG}. For the time being, we just assume that it is a quadratic function of the fields.
Note that the effective action~\eqref{eq:effective-action} can generally describe a quench because of the presence of the Heaviside step function in the second term. The field-independent factors $Z$, $K$ and $D$ account for possible renormalizations of the derivatives and of the Markovian noise, while the generic potential $\mathcal{U}(\phi)$ is a $\mathbb{Z}_2$-symmetric local polynomial of the order parameter $\phi$. For constructing the FRG equations, we consider the following cutoff function
\begin{equation}
\label{eq:Litim}
R(q)=K(k^2-q^2)\vartheta(k^2-q^2),
\end{equation}
which is known to minimise spurious effects introduced by the specific truncation ansatz of the effective action~\cite{Litim2000}.

The kernel $\Gamma^{(2)}+R\sigma$ appearing in Eq.~\eqref{eq:G-definition} --- which is obtained by deriving Eq.~\eqref{eq:effective-action} --- can be conveniently re-expressed by separating the field-independent part $G^{-1}_0$ (which receives contributions from the quadratic part of $\Gamma$ and from $\sigma R$) from the field-dependent part $V$, i.e., 
\begin{equation}
\label{eq:G0-V-definition}
\Gamma^{(2)}(x,x') + R(x,x') \sigma =G^{-1}_0(x,x') - V(x,x'),
\end{equation}
such that (see Eq.~\eqref{eq:G-definition})
\begin{equation}
\label{eq:G-G0-V}
G^{-1}(x,x') = G^{-1}_0(x,x') -V(x,x').
\end{equation}
Note that, since we assumed $\Gamma_0[\phi_0,\res_0]$ to be quadratic, its presence is completely encoded in the function $G^{-1}_0$. 
For the ansatz~\eqref{eq:effective-action}, the field-dependent part $V$ reads:
\begin{equation}
\label{eq:self-energy}
V(x,x') = V(x)\, \delta(x-x'),
\end{equation}
where the delta function $\delta(x-x')\equiv \delta(t-t')\delta^{(d)}(\rr-\rr')$ appears as a consequence of the locality in space and time of the potential $\mathcal{U}$, and where the function $V(x)$ is defined as
\begin{equation}
\label{eq:V-definition}
V(x) = -\vartheta(t-t_0)\begin{pmatrix}
\displaystyle{\res(x)\frac{\partial^3 \mathcal{U}}{\partial\phi^3}(x)} & \displaystyle{\frac{\partial^2 \mathcal{U}}{\partial \phi^2}(x)} \\[4mm]
\displaystyle{\frac{\partial^2 \mathcal{U}}{\partial \phi^2}(x)} & 0
\end{pmatrix}.
\end{equation}
The function $\vartheta$ in this expression of $V(x)$ appears as a consequence of the one in Eq.~\eqref{eq:effective-action}: as it will become clear below, the presence of $\vartheta$ allows one to encompass both the case of the quench and of a stationary state in the calculation of $G$ (see Sec.~\ref{sec:recovering-equilibrium}). 

Finally, in order to derive the RG equations for the couplings appearing in the effective action~\eqref{eq:effective-action},  one has to take the derivative with respect to $k$ on both sides of Eq.~\eqref{eq:vertexp-maintext} and, by using Eq.~\eqref{eq:wetterich}, one finds 
\begin{multline}
 \frac{\dd g_{n,l}}{\dd k} =\frac{\delta^{l+n}}{\delta\phi^n\delta\res^l} \frac{1}{2} \int_x  \text{tr}\left[ \vartheta(t-t_0)G(x,x)\frac{\dd R}{\dd k}\sigma\right]\biggr|_{\substack{\res =0 \\ \phi =\phi_m}} \\
 + \frac{\delta^{l+n+1} \Gamma}{\delta\phi^{n+1}\delta\res^l}\biggr|_{\substack{\res =0 \\ \phi =\phi_m}} \frac{\dd \phi_m}{\dd k},
\end{multline}
from which one can evaluate the flow equation for the couplings $g_{n,m}$, once the derivative of $\phi_m$ is calculated, where $\phi_m$ corresponds to the minimum of the potential $\mathcal{U}$.

\subsection{Functional renormalization group for a quench}
\label{sec:QFRG} 
In order to study the critical properties of the temperature quench described in Sec. \ref{sec:modelA}, we consider the effective action~\eqref{eq:effective-action}, in which one has still to specify the form of the boundary action $\Gamma_0$. The Gaussian probability distribution ~\eqref{eq:initial-probability}  of the initial condition can be effectively accounted for by taking
\begin{equation}
\label{eq:boundary-action}
\Gamma_0 = \int_\rr \left( -\frac{Z^2_0}{2\tau_0}\res_0^2 + Z_0\res_0\phi_0 + Z_0 h_0\res_0\right). 
\end{equation}
This form is uniquely fixed by requiring that it does not result in a violation of causality in the response functional~\footnote{In fact, one may be tempted to include in ${\Gamma_0}$ the term ${\propto \phi_0^2}$ appearing into the initial probability~\eqref{eq:initial-probability}: however, this would cause a violation of causality in the response functional, since each term in the action $\Gamma$ in Eq.~\eqref{eq:effective-action} must contain at least one response field ${\res}$ (see ,for instance, Ref.~\onlinecite{Tauberbook2014}). Instead, the term ${\propto \res_0^2}$ in Eq.~\eqref{eq:boundary-action} is allowed in this respect.}
and that it reproduces the Gaussian Green's functions~\eqref{eq:GR-gaussian} and~\eqref{eq:GK-gaussian}, see App.~\ref{app:dyson-initial}.
The factor $Z_0$ accounts for a possible renormalization of the initial response field $\res_0$: the way in which corrections to $Z_0$ are generated is discussed further below in this section. Note, in addition,  that the term $\propto \res_0^2$ can be regarded as a Gaussian noise located at the initial time $t_0$. 
The boundary action $\Gamma_0$ may in principle contain higher powers of $\phi_0$ and $\res_0$, and spatial and temporal derivatives of these fields: however, taking into account  their engineering dimension, one can argue~\cite{Janssen1989} that they are irrelevant in the renormalization-group sense, and therefore they have not been included here. The presence of a non-vanishing initial field $h_0$ induces a non-trivial evolution of the magnetization $M(t)$, but it does not generate new additional critical exponents (see Sec.~\ref{sec:modelA} and Ref.~\onlinecite{Janssen1989}), and therefore in the rest of this work we will assume $h_0 = 0$ without loss of generality.

In order to study the flow of the couplings of the effective action $\Gamma$ in Eq.~\eqref{eq:effective-action} from the FRG equation~\eqref{eq:wetterich} it is necessary to evaluate the matrix $G$ defined in Eq.~\eqref{eq:G-definition}. However, the presence of the boundary action given in Eq.~\eqref{eq:boundary-action} as well as the breaking of time-translational invariance in Eq.~\eqref{eq:effective-action} makes the calculation of $G(x,x')$ non-trivial, since now $G$ depends separately on the two times $t$ and $t'$. In order to overcome this difficulty, we notice that $G$ satisfies the following integral equation (see App.~\ref{app:Dyson} for a proof of this statement):
\begin{align}
\label{eq:Dyson-maintext}
G(x,x') 	 = G_0(x,x') + \int_y\, G_0(x,y)\, V(y) G(y,x'),
\end{align}
with $G_0$ and $V$ defined in Eq.~\eqref{eq:G0-V-definition}. The explicit form of $G_0$ can be evaluated by using the boundary action $\Gamma_0$ in Eq.~\eqref{eq:boundary-action} and it reads (see App.~\ref{app:dyson-initial} for details):
\begin{equation}
\label{eq:G0-definition}
G_0(t,t') = 
\begin{pmatrix}
G_{0C}(t,t') & G_{0R}(t,t') \\
G_{0R}(t',t) & 0
\end{pmatrix},
\end{equation}
where $G_{0R}$ and $G_{0C}$ are given by Eqs.~\eqref{eq:GR-gaussian} and Eqs.~\eqref{eq:GK-gaussian} (with $\Omega=1$ and $T$ replaced by $D$), respectively, with the dispersion relation $\omega_q$ replaced by the regularized one $\omega_{k,q}$, defined as:
\begin{equation}
\label{eq:modified-dispersion}
\omega_{k,q} \equiv Kq^2 + \tau + K(k^2-q^2)\vartheta(k^2-q^2).
\end{equation}
Equation~\eqref{eq:Dyson-maintext} can then be solved iteratively and, once its  solution has been replaced into the FRG equation~\eqref{eq:wetterich}, the latter can be cast in the form (see, e.g., Eqs.~\eqref{eq:wetterich-series} and~\eqref{eq:DeltaGamma-n}):
\begin{equation}
\label{eq:wetterich-series-maintext}
\frac{\dd \Gamma}{\dd k} = \sum_{n=1}^{+\infty} \Delta \Gamma_n, 
\end{equation}  
where the functions $\Delta \Gamma_n$ are defined as (see App.~\ref{app:Dyson} for details)
\begin{multline}
\Delta \Gamma_n  = \frac{1}{2} \int_{x,y_1\dots y_n}  \text{tr}\bigg[ G_0(x,y_1)V(y_1)G_0(y_1,y_2) \\
\times \dots \times V(y_n)G_0(y_n,x)\frac{\dd R}{\dd k}\sigma\bigg]. \label{eq:DeltaGamma-n-maintext}
\end{multline}
As discussed in App.~\ref{app:Dyson}, the FRG equation in the form of Eq.~\eqref{eq:wetterich-series-maintext} is the most convenient one for calculations when, as in the present case, time-translational invariance is broken and  therefore it cannot be further simplified by expressing it in the Fourier frequency space.

For simplicity, let us assume that the potential $\mathcal{U}$  in Eq.~\eqref{eq:effective-action} is quartic in the field $\phi$, i.e.,
\begin{equation}
\label{eq:potential-quartic}
\mathcal{U}(\phi) = \frac{\tau}{2}\phi^2 + \frac{g}{4!}\phi^4,
\end{equation}
such that, from Eq.~\eqref{eq:V-definition}, the field-dependent function $V(x)$ reads
\begin{equation}
\label{eq:self-energy-quartic}
V(x) = -\vartheta(t-t_0)\, g
\begin{pmatrix}
\res\phi 				& \phi^2/2 \\
\phi^2/2 	&	 0
\end{pmatrix}.
\end{equation}
Accordingly, since this $V$ appears $n$ times in the convolution~\eqref{eq:DeltaGamma-n-maintext} which defines $\Delta \Gamma_n$ on the r.h.s.~of Eq.~\eqref{eq:wetterich-series-maintext}, it follows that $\Delta\Gamma_n$ contains products of $2n$ possibly different fields. Because of the ansatz~\eqref{eq:effective-action}  also the l.h.s.~of Eq.~\eqref{eq:wetterich-series-maintext} is a polynomial of the fields, and therefore each term on the l.h.s.~is  uniquely matched by a term of the expansion on the r.h.s. Accordingly, in order to derive the RG equation for the coupling of a term involving a product of $2n$ fields, it is sufficient to evaluate the corresponding $\Delta \Gamma_n$. 
Note that this line of argument applies also to the time-translational invariant case, and, moreover, it  can be easily generalized to the case in which the potential contains powers of $\phi$  of higher order than those in Eq. \eqref{eq:potential-quartic}. 
        
\section{Truncation for $\phi_m=0$}
\label{sec:truncation-symmetric}
In this section we discuss the derivation of the RG equations from the ansatz~\eqref{eq:effective-action} with the quartic potential $\mathcal{U}$ introduced in Eq.~\eqref{eq:potential-quartic}. 
Considering this simple case allows us to detail how the boundary action~\eqref{eq:boundary-action} is renormalized by the post-quench interaction.
Since this ansatz corresponds to a local potential approximation~\cite{Berges2002, Delamotte2007,Canet2005}, the anomalous dimensions of the derivative terms ($K,Z$) and of the Markovian noise strength $D$ vanish, and therefore in the following we set, for simplicity, $K = Z =1$. 
The only non-irrelevant terms which are renormalized within this scheme are those proportional to quadratic and quartic powers of the fields $\phi$ and $\res$, i.e., those associated with the post-quench parameter $\tau$, the boundary field renormalization $Z_0$ and the coupling $g$. 
As discussed in Sec.~\ref{sec:QFRG}, the renormalization of the quadratic terms is determined by the contribution $\Delta \Gamma_1$ appearing on the r.h.s.~of Eq.~\eqref{eq:wetterich-series-maintext}, while the renormalization of the quartic one by the contribution $\Delta \Gamma_2$.

\subsection{Derivation of the RG equations}
\label{sec:derivation-RG-equations}
Let us now consider Eq.~\eqref{eq:wetterich-series-maintext} and focus on the term $\Delta \Gamma_1$, as defined in Eq.~\eqref{eq:DeltaGamma-n-maintext}. A simple calculation renders (see App.~\ref{app:evaluation-DGamma} for details)
\begin{align}
\label{eq:dGamma-1}
\Delta \Gamma_1
& = -k^{d+1}\frac{a_d}{d}\frac{g D}{\omega_k^2} \int_\rr \int_{t_0}^{+\infty} \dd t\, \res(t,\rr)\phi(t,\rr) \nonumber \\
& \qquad \qquad \qquad \times \left[ 1 - f_\tau(t-t_0)\right], 
\end{align}
where $a_d = 2/[\Gamma(d/2) (4\pi)^{d/2}]$, with $d$ the spatial dimensionality of the system and $\Gamma(x)$ the gamma function. The integration over the intermediate time variable in Eq.~\eqref{eq:DeltaGamma-n-maintext}  for $n=1$ generates, within the square brackets in the integrand of Eq.~\eqref{eq:dGamma-1}, one term which is independent of time and one which depends on it via the function $f_\tau(t-t_0)$, defined as
\begin{equation}
\label{eq:f-definition}
f_\tau(t) = \ee^{-2\omega_k t} \left[1+2\omega_kt\left(1-\frac{\omega_k}{D\tau_0}\right)\right],
\end{equation}
where $\omega_k \equiv \omega_{q=k}$ or, equivalently, $\omega_k\equiv \omega_{k,q=k}$ (see Eq.~\eqref{eq:modified-dispersion}).
Since $f_\tau(t)$ vanishes exponentially fast upon increasing the time $t$, its contribution to the renormalization of the time-independent parameter $\tau$ can be neglected~\footnote{In this calculation and in those which follow, we always take first the limit ${t \to \infty}$, and then the limit ${k \to 0}$.}. 
Accordingly, the flow equation for $\tau$ can be simply obtained by comparing the l.h.s.~of Eq.~\eqref{eq:wetterich-series-maintext} with Eq.~\eqref{eq:dGamma-1}, where we introduced the potential~\eqref{eq:potential-quartic} in the truncated action~\eqref{eq:effective-action}; this yileds 
\begin{equation}
\label{eq:r-eq}
\frac{\dd \tau}{\dd k} = -k^{d+1}\frac{a_d}{d}\frac{g D}{(k^2 + \tau)^2}.
\end{equation}
At short times, instead, the function $f_\tau(t)$ singles out contributions containing fields of the temporal boundary, thus renormalizing the boundary action $\Gamma_0$ introduced in Eq.~\eqref{eq:boundary-action}. In fact, the formal identity
\begin{equation}
\int_{t_0}^{+\infty} \dd t\, g(t) \ee^{-c(t-t_0)} = \sum_{n=0}^{+\infty} \frac{1}{c^{n+1}} \frac{\dd^n g}{\dd t^n}\biggr|_{t=t_0},
\end{equation}
with $c>0 $ and $g(t)$ an arbitrary smooth function, can be used in order to express  the part of the integral involving $f_\tau(t)$ on the r.h.s.~of Eq.~\eqref{eq:dGamma-1} as
\begin{multline}
\label{eq:boundary-generation}
\int_{t_0}^{+\infty}\dd t\, \res(t)\phi(t) f_\tau(t-t_0) = \\
=\sum_{n=0}^{+\infty} \frac{c_{n,k}(\tau_0)}{(2\omega_k)^{n+1}}\,  Z_{0,n} \frac{\dd^n }{\dd t^n} \left[\res(t)\phi(t)\right]\biggr|_{t=t_0},
\end{multline}
with 
\begin{equation}
c_{n,k}(\tau_0) \equiv (n+2) - \frac{(n+1) \omega_k}{D\tau_0}.
\end{equation}
Accordingly, the time-dependent part in the integrand of Eq.~\eqref{eq:dGamma-1} generates an infinite series of operators contributing to the boundary action $\Gamma_0$. For future convenience, we introduced in Eq.~\eqref{eq:boundary-generation} additional numerical factors $Z_{0,n}$, which account for  possible renormalization of the boundary operators and which  equal one in the non-renormalized theory. 
Most of the terms in the sum~\eqref{eq:boundary-generation} renormalize irrelevant operators which were not included into the original ansatz~\eqref{eq:boundary-action} for the boundary action, and therefore one can neglect them. The only non-irrelevant term corresponds to $n=0$ in Eq.~\eqref{eq:boundary-generation}: by inserting the boundary action $\Gamma_0$ (see Eq.~\eqref{eq:boundary-action}) into the l.h.s.~of Eq.~\eqref{eq:wetterich-series-maintext}, and by combining it with Eqs.~\eqref{eq:dGamma-1} and~\eqref{eq:boundary-generation}, one finds the flow equation for $Z_0 \equiv Z_{0,0}$, i.e., 
\begin{equation}
\label{eq:Z0-eq}
\frac{\dd Z_0}{\dd k}  =  k^{d+1}\frac{a_d}{d}\frac{g D}{(k^2 + \tau)^3}\left[1 - \frac{k^2+ \tau}{2 D\tau_0}\right]\, Z_0.
\end{equation}
We consider now the renormalization of the quartic term, which can be read off from $\Delta \Gamma_2$. A simple calculation renders (see App.~\ref{app:evaluation-DGamma} for details)
\begin{align}
\label{eq:dGamma-2}
&\Delta \Gamma_2  = \nonumber \\ 
& = \frac{3}{2}k^{d+1}\frac{a_d}{d}\frac{g^2D^2}{\omega^4_k} \int_{\rr}\int_{t_0}^{+\infty} \dd t\,\res^2(t)\phi^2(t) \left[1 - f_D(t-t_0)\right]  \nonumber \\
& + k^{d+1}\frac{a_d}{d}\frac{g^2D}{\omega_k^3} \int_{\rr}\int_{t_0}^{+\infty} \dd t\, \res(t')\phi^3(t)\left[1 - f_g(t-t_0)\right],
\end{align}
where  $f_g$ and $f_D$, given in, cf., Eqs.~\eqref{eq:fD} and~\eqref{eq:fg}, respectively, decay exponentially upon increasing the time $t$, and therefore they do not contribute to the renormalization of the couplings at long times. Note that the integration produces a term proportional to $\res^2\phi^2$ in Eq.~\eqref{eq:dGamma-2}: however, this operator is irrelevant for $d>2$ and  it can  be neglected, since our truncation includes only relevant couplings. On the other hand, the term proportional to $\res\phi^3$ in Eq.~\eqref{eq:dGamma-2}  renormalizes the relevant coupling $g$ and, comparing Eq.~\eqref{eq:dGamma-2} with the l.h.s.~of Eq.~\eqref{eq:wetterich-series-maintext} after using the ansatz~\eqref{eq:effective-action} for $\Gamma$ with the potential~\eqref{eq:potential-quartic}, one finds the flow equation for $g$: 
\begin{equation}
\label{eq:g-eq}
\frac{\dd g}{\dd k } = 6k^{d+1}\frac{a_d}{d}\frac{g^2D}{(k^2 + \tau)^3}.
\end{equation}

\subsection{Flow equations}
\label{sec:flow-equations}
In order to study the flow of $\tau$ and $g$ prescribed by Eqs.~\eqref{eq:r-eq} and~\eqref{eq:g-eq}, it is convenient to introduce the dimensionless quantities $\widetilde{\tau}=\tau/k^2$, and $\widetilde{g}= g D  k^{d-4}a_d/d$. The corresponding flow equations follow from Eqs.~\eqref{eq:r-eq} and~\eqref{eq:g-eq}:
\begin{align}
k\frac{\dd \widetilde{\tau}}{\dd k}	& = -2\widetilde{\tau}  - \frac{\widetilde{g}}{(1+\widetilde{\tau})^2}, \label{eq:r-eq-dimensionless}\\
k\frac{\dd \widetilde{g}}{\dd k} 	& = \widetilde{g}\left[ -\epsilon + 6\frac{\widetilde{g}}{(1+\widetilde{\tau})^3} \right], \label{eq:g-eq-dimensionless}
\end{align}
where $\epsilon = 4-d$. These equations describe the RG flow of the couplings in the equilibrium state which is asymptotically reached by the system at long times. Accordingly, they are independent of both $Z_0$ and $\tau_0$:  the relaxational nature of model A erases the information about the initial state in the long time. 
Since the final state corresponds to an equilibrated system, the equations for $\widetilde{\tau}$ and $\widetilde{g}$ must result into the same critical exponents as in the equilibrium Ising universality class~\cite{Goldenfeldbook,Mabook,Tauberbook2014}. This can be seen, for instance, by comparing Eqs.~\eqref{eq:r-eq-dimensionless} and~\eqref{eq:g-eq-dimensionless} (at leading order in $\epsilon$) with the results obtained within the perturbative RG at one loop in the equilibrium theory~\cite{Tauberbook2014}. Note that Eqs.~\eqref{eq:r-eq-dimensionless} and~\eqref{eq:g-eq-dimensionless} do not have the same form as the corresponding equations derived within perturbative RG, as they are obtained within a different renormalization scheme; nevertheless, they provide the same critical exponents, as discussed further below. 
 
Equations~\eqref{eq:r-eq-dimensionless} and~\eqref{eq:g-eq-dimensionless} admit two fixed points: the Gaussian one $(\widetilde{\tau}_\text{G}^*,\widetilde{g}_\text{G}^*) = (0,0)$ and the Wilson-Fisher one, which at leading order in $\epsilon$, reads $(\widetilde{\tau}^*_\text{WF},\widetilde{g}^*_\text{WF})= (-\epsilon/12, \epsilon/6) + O(\epsilon^2)$ (in general we will denote by the superscript $^*$ any quantity which is evaluated at a fixed point). By linearizing Eqs.~\eqref{eq:r-eq-dimensionless} and~\eqref{eq:g-eq-dimensionless} around these fixed points, one finds that the Gaussian one is stable only for $\epsilon <0$, while the Wilson-Fisher fixed point is stable only for $\epsilon >0$. The latter has an unstable direction, and from the inverse of the negative eigenvalue of the associated stability matrix, one derives the critical exponent $\nu$, which reads $\nu=1/2+\epsilon/12 + O(\epsilon^2)$, which is the same as in equilibrium~\cite{Goldenfeldbook,Mabook,Tauberbook2014}.
As mentioned at the beginning of this section, the ansatz~\eqref{eq:potential-quartic} for the potential does not allow for a renormalization of the time and spatial derivatives in the effective action~\eqref{eq:effective-action}. Accordingly, the anomalous dimension $\eta$ and the dynamical critical exponent $z$ are equal to their Gaussian values $\eta = 0$ and $z =2$.

Let us now focus on the renormalization of the terms in the boundary action $\Gamma_0$ in Eq.~\eqref{eq:boundary-action}. From Eq.~\eqref{eq:Z0-eq}, we define the anomalous dimension $\eta_0$ of the response field $\res_0$ at initial time as
\begin{equation}
\label{eq:Z0-eq-dimensionless}
\eta_0 \equiv -\frac{k}{Z_0}\frac{\dd Z_0}{\dd k} = -\frac{\widetilde{g}}{(1+\widetilde{\tau})^3} \left(1-\frac{1+\widetilde{\tau}}{2\widetilde{\tau}_0}\right), 
\end{equation}
where we introduced the rescaled pre-quench parameter $\widetilde{\tau_0} = \tau_0/k^2$ and we used Eq.~\eqref{eq:Z0-eq}. Since $\tau_0$ does not receive any correction from the renormalization, its flow equation is simply determined by its canonical dimension and thus
\begin{equation}
\label{eq:tau0-eq-dimensionless}
k\frac{\dd \widetilde{\tau}_0}{\dd k}	= -2\widetilde{\tau}_0.
\end{equation}
Accordingly, $\widetilde{\tau}_0$ has only one stable fixed point $\widetilde{\tau}^*_0 = +\infty$, in the infrared regime (i.e., for $k \to 0$). Close to this fixed point, any possible term in the boundary action $\Gamma_0$ (except for $\res_0\phi_0$) is irrelevant for $d>2$, and therefore the ansatz~\eqref{eq:boundary-action} is consistent. Note that the r.h.s.~of Eq.~\eqref{eq:Z0-eq-dimensionless} diverges at the unstable fixed point $\widetilde{\tau}^*_0 =0$: this is expected since $\tau_0 = 0$ is unphysical~\cite{Janssen1989} for the initial probability in Eq.~\eqref{eq:initial-probability} and hence for the ansatz in Eq.~\eqref{eq:boundary-action}, as it would correspond to a non-normalizable probability.

The value $\eta^*_0$ of the anomalous dimension $\eta_0$ of the initial response field at the Wilson-Fisher fixed point can be straightforwardly derived by substituting in Eq.~\eqref{eq:Z0-eq-dimensionless} the fixed-point values $\widetilde{\tau}^*_\text{WF}$ and $\widetilde{g}^*_\text{WF}$ of the couplings, obtaining $\eta^*_0 = -\epsilon/6$. 
The initial-slip exponent $\theta$ is then defined as~\cite{Janssen1989,Calabrese2005}:
\begin{equation}
\label{eq:theta-definition}
\theta = -\frac{\eta_0^*}{z},
\end{equation}
and therefore, in the present case, it takes the value 
\begin{equation}
\theta = -\frac{\eta^*_0}{2} = \frac{\epsilon}{12},
\end{equation}
which agrees up to first order in $\epsilon$, with the expression 
\begin{equation}
\label{eq:theta-Janssen}
\theta = \frac{\epsilon}{12}\left[ 1  + \epsilon \left(\frac{8}{27}+\frac{2\log 2}{3}\right)\right] +O(\epsilon^3),
\end{equation}
obtained in Ref.~\onlinecite{Janssen1989}.

\subsection{Comparison with equilibrium dynamics}
\label{sec:recovering-equilibrium}
In this section, we show how one can recover the flow equations for the equilibrium case in the limit $t_0\to-\infty$. First of all, we note that in the expressions for $\Delta\Gamma_1$ and $\Delta\Gamma_2$ given in Eqs.~\eqref{eq:dGamma-1} and~\eqref{eq:dGamma-2}, respectively, the only dependence on $t_0$ occurs in the lower limit of the integration domain of the integrals on $t$ and in the functions $f_\tau(t-t_0)$, $f_g(t-t_0)$, and $f_D(t-t_0)$. For $t_0\to-\infty$ these functions vanish exponentially fast (see Eqs.~\eqref{eq:f-definition}, \eqref{eq:fg} and~\eqref{eq:fD}) and Eqs.~\eqref{eq:dGamma-1} and~\eqref{eq:dGamma-2} read
\begin{subequations}
\label{eq:dGamma-eq}
\begin{align}
\Delta \Gamma^\text{eq}_1 & = -k^{d+1}\frac{a_d}{d}\frac{g D}{\omega_k^2} \int_x \res(x)\phi(x) \label{eq:dGamma1-eq},\\
\Delta \Gamma^\text{eq}_2 & =  k^{d+1}\frac{a_d}{d}\frac{g^2D}{\omega^3_k} \int_x \left[ \frac{3D}{2\omega_k}\res^2(x)\phi^2(x) + \res(x)\phi^3(x)\right],\label{eq:dGamma2-eq}
\end{align}
\end{subequations}
with $x \equiv (\rr,t)$ and $\int_x \equiv \int \dd^dr\int_{-\infty}^{+\infty} \dd t$.

Alternatively, one could have taken the limit $t_0\to -\infty$ from the outset, i.e., before evaluating $\Delta\Gamma_1$ and $\Delta \Gamma_2$: in this case one simply needs to replace $\vartheta(t-t_0)$ with its limiting value $1$ in Eqs.~\eqref{eq:wetterich} and~\eqref{eq:V-definition}, while $G_{0R}$ is modified inasmuch $G_{0C}$ becomes time-translational invariant as $t_0 \to -\infty$ (see Eqs.~\eqref{eq:GK-gaussian} and~\eqref{eq:G0-definition}). This gives rise again to Eqs.~\eqref{eq:dGamma-eq}, since the operations of taking the limit $t_0 \to -\infty$ and of calculating the integrals over time (and momenta) on the r.h.s.~of Eq.~\eqref{eq:wetterich-series-maintext} do commute (because all the time integrals  are convergent due to the decreasing exponentials in $G_{0R}$ and $G_{0C}$). 

Taking the limit $t_0\to -\infty$ in the action~\eqref{eq:effective-action} just corresponds to consider the equilibrium, time-translational invariant theory~\cite{Tauberbook2014},  
and therefore one concludes that Eqs.~\eqref{eq:dGamma-eq} give rise to the equilibrium flow equations. Since the flow equations~\eqref{eq:r-eq-dimensionless} and~\eqref{eq:g-eq-dimensionless} can also be derived from Eqs.~\eqref{eq:dGamma-eq}, they thus represent the equilibrium ones: this is an expected result, since the relaxational nature of model A leads the system to its equilibrium state (yet for asymptotically long times at the critical point), regardless of the quench protocol~\cite{Janssen1989}.  

\section{Truncation for $\phi_m \neq 0$}
\label{sec:truncation-ordered}
In this section, we discuss the results of a different, improved ansatz for the potential $\mathcal{U}$ in the effective action~\eqref{eq:effective-action}, namely
\begin{equation}
\label{eq:potential-ordered}
\mathcal{U} = \frac{g}{4!}(\phi^2-\phi_m^2)^2+\frac{\lambda}{6!}(\phi^2-\phi_m^2)^3, 
\end{equation}
the flow of which is derived in App.~\ref{app:beta_ordered}.
This potential differs from the one considered in Eq.~\eqref{eq:potential-quartic} in two respects. First, it corresponds to an expansion around a finite homogeneous value $\phi_m$: this choice has the leverage to capture the leading divergences of two-loops corrections in a calculation which is technically carried at one-loop, as typical of  background field methods (see, e.g.,  Refs.~\onlinecite{Berges2002, Delamotte2007,Canet2007,Sieberer2014}), and thus it allows us to calculate, for instance, the renormalization of the factors $Z$, $K$ and $D$.
In fact, the presence of a background field, $\phi_m$, reduces two-loop diagrams to one-loop ones in which an internal classical line (corresponding to a correlation function, $G_C$) has been replaced by the insertion of two expectation values $\phi_m$  (straight lines stand for the field $\phi$, curved lines for the response field $\tilde{\phi}$; see, e.g., Ref.~\onlinecite{Tauberbook2014}). For instance, the renormalization of $Z$ and $K$ comes from the diagram
\begin{equation}
\begin{tikzpicture}[baseline={([yshift=-.5ex]current bounding box.center)}]
\coordinate[label={[label distance=-1]above:$G_C$}] (o) at (0,0);
\coordinate[label={[label distance=3.5]above left:$g$}] (ol) at (-0.5,0);
\coordinate[label={[label distance=3.5]above right:$g$}] (or) at (0.5,0);
\coordinate[] (oll) at (-1.25,0);
\coordinate[] (orr) at (1.25,0);
\coordinate[label={[label distance=-1]above:$G_C$}] (on)  at (0,0.5);
\coordinate[label={[label distance=0]below:$G_R$}] (on)  at (0,-0.5);
\draw[thick] (or) -- (ol);
\draw[res] (or) arc (0:-90:0.5);
\draw[thick] (or) arc (0:270:0.5);
\draw[res] (ol) node[circle,fill,inner sep=1pt]{} -- (oll);
\draw[thick] (or) node[circle,fill,inner sep=1pt]{} -- (orr);
\end{tikzpicture}
\quad \Longrightarrow
\begin{tikzpicture}[baseline={([yshift=-.5ex]current bounding box.center)}]
\coordinate[] (o) at (0,0);
\coordinate[label={[label distance=3.5]left:$g$}] (ol) at (-0.5,0);
\coordinate[label={[label distance=3.5]right:$g$}] (or) at (0.5,0);
\coordinate[label={[label distance=1]above left:$\phi_m$}] (no) at (-1.25,0.5);
\coordinate[label={[label distance=1]above right:$\phi_m$}] (ne) at (1.25, 0.5);
\coordinate[] (so) at (-1.25,-0.5);
\coordinate[] (se) at (1.25,-0.5);
\coordinate[label={[label distance=-1]above:$G_C$}] (on)  at (0,0.5);
\coordinate[label={[label distance=0]below:$G_R$}] (on)  at (0,-0.5);
\draw[res] (or) arc (0:-90:0.5);
\draw[thick] (or) arc (0:270:0.5);
\draw[thick] (ol) node[circle,fill,inner sep=1pt]{} -- (no) node[draw,cross out,rotate=60]{};
\draw[res] (so) -- (ol);
\draw[thick] (ne) node[draw,cross out,rotate=30]{} -- (or);
\draw[thick] (or) node[circle,fill,inner sep=1pt]{} -- (se);
\end{tikzpicture}, 
\end{equation}
while the renormalization of the noise strength $D$ comes from the diagram:
\begin{equation}
\begin{tikzpicture}[baseline={([yshift=-.5ex]current bounding box.center)}]
\coordinate[label={[label distance=-1]above:$G_C$}] (o) at (0,0);
\coordinate[label={[label distance=3.5]above left:$g$}] (ol) at (-0.5,0);
\coordinate[label={[label distance=3.5]above right:$g$}] (or) at (0.5,0);
\coordinate[] (oll) at (-1.25,0);
\coordinate[] (orr) at (1.25,0);
\coordinate[label={[label distance=-1]above:$G_C$}] (on)  at (0,0.5);
\coordinate[label={[label distance=0]below:$G_C$}] (on)  at (0,-0.5);
\draw[thick] (or) -- (ol);
\draw[thick] (or) arc (0:-90:0.5);
\draw[thick] (or) arc (0:270:0.5);
\draw[res] (ol) node[circle,fill,inner sep=1pt]{} -- (oll);
\draw[res] (or) node[circle,fill,inner sep=1pt]{} -- (orr);
\end{tikzpicture}
\quad \Longrightarrow
\begin{tikzpicture}[baseline={([yshift=-.5ex]current bounding box.center)}]
\coordinate[] (o) at (0,0);
\coordinate[label={[label distance=3.5]left:$g$}] (ol) at (-0.5,0);
\coordinate[label={[label distance=3.5]right:$g$}] (or) at (0.5,0);
\coordinate[label={[label distance=1]above left:$\phi_m$}] (no) at (-1.25,0.5);
\coordinate[label={[label distance=1]above right:$\phi_m$}] (ne) at (1.25, 0.5);
\coordinate[] (so) at (-1.25,-0.5);
\coordinate[] (se) at (1.25,-0.5);
\coordinate[label={[label distance=-1]above:$G_C$}] (on)  at (0,0.5);
\coordinate[label={[label distance=0]below:$G_C$}] (on)  at (0,-0.5);
\draw[thick] (or) arc (0:-90:0.5);
\draw[thick] (or) arc (0:270:0.5);
\draw[thick] (ol) node[circle,fill,inner sep=1pt]{} -- (no) node[draw,cross out,rotate=60]{};
\draw[res] (so) -- (ol);
\draw[thick] (ne) node[draw,cross out,rotate=30]{} -- (or);
\draw[res] (or) node[circle,fill,inner sep=1pt]{} -- (se);
\end{tikzpicture}.
\end{equation}
Second, we added a sextic interaction, which is marginal for $d=3$ and therefore it is expected to contribute with sizeable corrections to the value of the critical exponents only upon approaching $d = 3$. In fact, the effective action~\eqref{eq:effective-action} with the potential~\eqref{eq:potential-ordered} contains all the non-irrelevant operators in $d=3$.
As anticipated, this ansatz allows the renormalization of the time and spatial derivatives terms and of the Markovian noise, i.e., of the coefficients $K$, $Z$ and $D$ in Eq.~\eqref{eq:effective-action}, which therefore will be reinstated in the following analysis. The flow equations for these coefficients can be conveniently expressed in terms of the corresponding anomalous dimensions $\eta_D, \eta_Z$ and $\eta_K$, defined as:
\begin{equation}
\label{eq:anomalous-dimensions}
\eta_D \equiv -\frac{k}{D}\frac{\dd D}{\dd k}, \quad  \eta_Z \equiv -\frac{k}{Z}\frac{\dd Z}{\dd k}, \quad  \eta_K \equiv -\frac{k}{K}\frac{\dd K}{\dd k}.
\end{equation}
The calculation of $\eta_D$, $\eta_Z$ and $\eta_K$  is detailed, respectively, in Apps.~\ref{app:anomalousD}, \ref{app:anomalousZ} and~\ref{app:anomalousK}.

The somewhat lengthy flow equations of the corresponding dimensionless couplings
\begin{equation}
\label{eq:couplings-dimensionless}
\widetilde{m}=\frac{1}{3}\frac{\phi^2_m g}{Kk^2}, \quad \widetilde{g}=\frac{a_d}{d}\frac{D}{ZK^2}\frac{g}{k^{4-d}}, \quad \widetilde{\lambda}=\frac{a_d}{d}\frac{D^2}
{Z^2K^3} \frac{\lambda}{k^{6-2d}}
\end{equation}
and of the anomalous dimensions $\eta_{D,Z,K}$ is reported in Eqs.~\eqref{eq:m-RG}-\eqref{eq:eta0-RG} of App.~\ref{app:FRGbetafunct}.
First of all, we note that $\eta_D = \eta_Z$: this a consequence of detailed balance~\cite{Tauberbook2014,Canet2007,Sieberer2014}, which characterizes the equilibrium dynamics of model A. In fact, while the short-time dynamics after the quench violates detailed balance inasmuch time-translational invariance is broken, in the long-time limit (in which the flow equations are valid) detailed balance is restored. 

The fixed points $(\widetilde{m}^*,\widetilde{g}^*,\widetilde{\lambda}^*)$ of Eqs.~\eqref{eq:m-RG}-\eqref{eq:etaK-RG} can be determined numerically (see App.~\ref{app:FRGbetafunct} for details) and they can be used in order to calculate the anomalous dimension $\eta$ and the dynamical critical exponent $z$ as
\begin{equation}
\eta = \eta_K^*, \qquad z = 2- \eta^*_K + \eta_Z^*.
\end{equation}
The critical exponent $\nu$ can be determined after linearizing the flow equations around the fixed point, as the inverse of the negative eigenvalue of the stability matrix (see App. \ref{app:FRGbetafunct}) .

As a consistency check, we compare our values $\nu=0.64$, $\eta=0.11$ and $z =2.05$ in $d=3$ with the ones determined in Ref.~\onlinecite{Canet2007} for the equilibrium dynamics of model A within the same truncation ansatz for the effective action $\Gamma$ as the one employed here, i.e., $\nu=0.65$, $\eta=0.11$ and $z =2.05$, finding very good agreement. For completeness, we also report the Monte Carlo estimates (see Ref.~\onlinecite{Canet2007} for a summary), given by $\nu_\text{MC} =0.6297(5)$, $\eta_\text{MC} =0.0362(8)$, and $z_\text{MC} = 2.055(10)$.

\begin{figure}[t!]\centering
\includegraphics[width=8.5cm]
{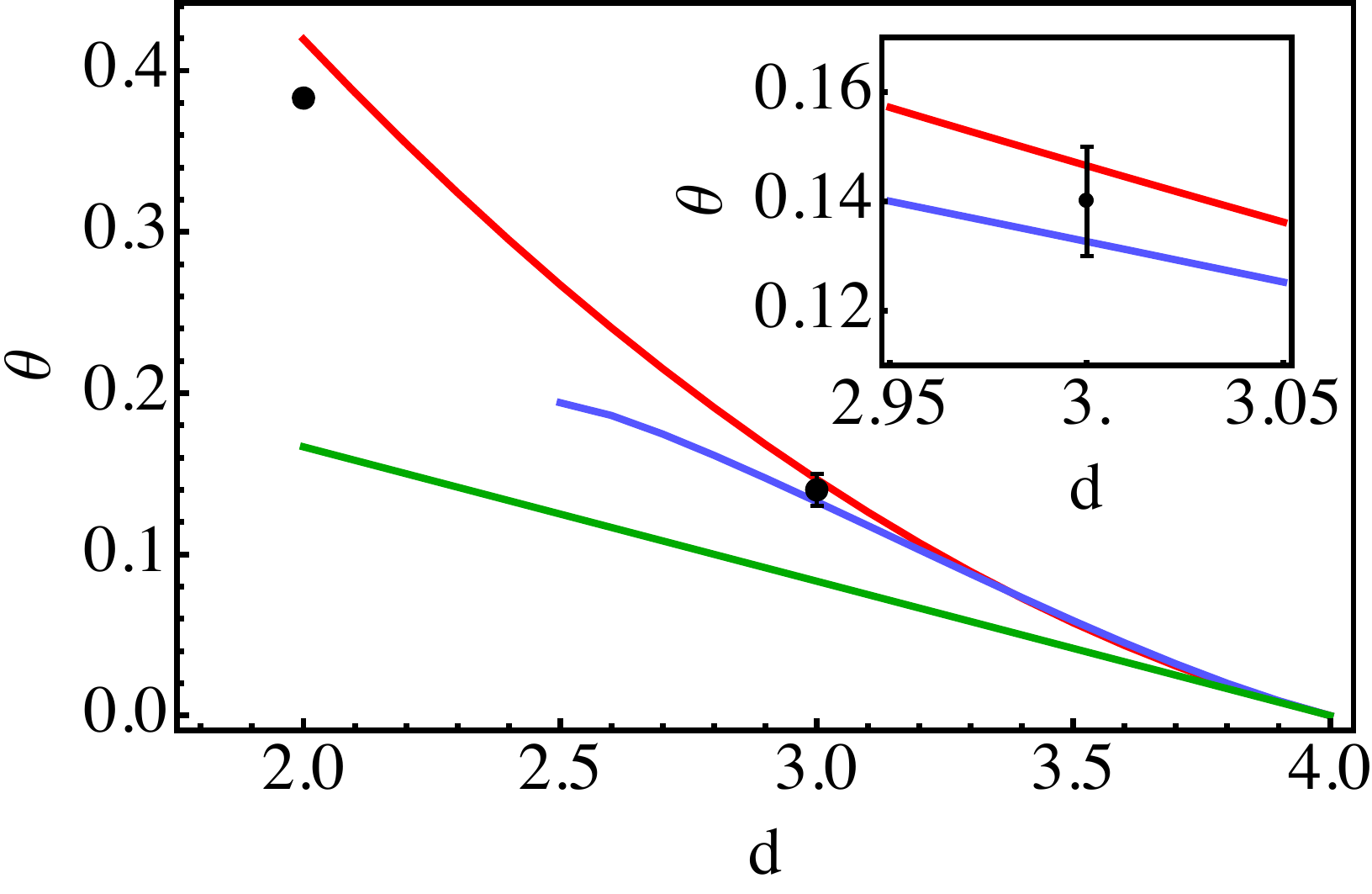}
 \caption{(Color online). Main plot: initial-slip exponent $\theta$ as a function of the spatial dimensionality $d$, evaluated from the FRG discussed here (blue, central line) and from the $\epsilon$-expansion to first (green, lower line) and second (red upper line) order in $\epsilon=4-d$ provided in Eq.~\eqref{eq:theta-Janssen}. The value of $\theta$ obtained from numerical Monte Carlo simulations are indicated for $d = 2$ and $3$ (symbols with error bars). For $d=2$, the error bars are within the symbol size. Inset: magnification of the main plot for $d\simeq 3$.}
\label{fig:fig2}
\end{figure}

In Fig.~\ref{fig:fig2}, we compare the values of $\theta$ obtained from Eq.~\eqref{eq:theta-definition} on the basis of the present analysis (blue line), and of the first- (green line) and second-order (red line) $\epsilon$-expansion of Ref.~\onlinecite{Janssen1989} reported in Eq.~\eqref{eq:theta-Janssen}, as a function of the spatial dimensionality $d$. 
The first-order term in the $\epsilon$-expansion is accurate only for spatial dimensionality $d$ close to $d = 4$, while the second-order contribution provides  sizeable corrections at smaller values of $d$. 
Our results are in remarkable agreement with the latter expansion for $d \gtrsim 3.2$, while increasing discrepancies emerge at smaller values of $d$. 
In particular, for $d \leq 3$ additional stable fixed points appears in the solution of  Eqs.~\eqref{eq:m-RG}-\eqref{eq:etaK-RG} beyond the Wilson-Fisher one, while for $d \leq 2.5$ the latter disappears. This is not surprising, since for $d \leq 3$ new non-irrelevant terms are allowed, and therefore the potential in Eq.~\eqref{eq:potential-ordered} is no longer an appropriate ansatz and additional terms have to be introduced.
In particular, the number of non-irrelevant operators diverges as $d$ approaches $2$: one should indeed recall that in $d = 2$ any term of the form $\res \phi^{2n+1}$, with positive integer $n$, is relevant in the RG sense and therefore the correct truncation for the effective action requires considering a full functional ansatz for the potential, beyond the polynomial expansion used in this work. In Refs.~\onlinecite{Delamotte2007,Canet2007} it is shown how to deal with this issue within the standard approach to FRG.

For comparison, we report in Fig.~\ref{fig:fig2} also the two values of $\theta$ obtained from Monte Carlo simulations (see, e.g., the summary in Ref.~\onlinecite{Calabrese2005}) in $d = 2$ and $d = 3$ (symbols). Remarkably, the predictions of both FRG and $\epsilon$-expansion are compatible (within error bars) with the numerical estimate in $d=3$, where the ansatz for the potential~\eqref{eq:potential-ordered} is reliable, while the FRG predicts a smaller value compared to the one predicted by the $\epsilon$-expansion. For $d=2$, instead, our ansatz~\eqref{eq:potential-ordered} is unable to provide reliable predictions for the reasons reported above, while the $\epsilon$-expansion still provides an unexpectedly accurate estimat, yet outside the error bars of the best available numerical estimate $\theta = 0.383(3)$.  

%
%
%

\section{Conclusions and perspectives} 
\label{sec:concls}
In this work we generalized the functional renormalization group (FRG) scheme in order to describe the universal dynamical behaviour emerging at short times in a classical statistical system after a temperature quench to its critical point. 
Specifically, we focused on the relaxational dynamics described by the model A~\cite{Tauberbook2014} for a scalar order parameter and a Landau-Ginzburg effective Hamiltonian, and we evaluated the initial-slip exponent $\theta$, which controls the universal scaling of correlation functions and magnetization after the quench within the Ising universality class with spin-flip (Glauber) dynamics. The value of $\theta$ is found to be in good agreement with the one obtained via an $\epsilon$-expansion and numerical simulations in $d=3$.  
Our prediction for $\theta$ can be systematically improved by using a more refined ansatz for the effective action, taking advantage of the existing FRG schemes for equilibrium systems~\cite{Berges2002,Delamotte2007}.

The approach developed in this work can be extended to different static universality classes, such as $O(N)$ and Potts models, or to different dynamics, e.g., with conserved quantities~\cite{Oerding1993, Oerding1993b}. In addition, it can also be used in order to study equilibrium phase transitions in systems with a spatial boundary, whose description is formally similar to the case of a quench~\cite{Diehl1997}, and possibly also their non-equilibrium dynamics~\cite{Marcuzzi2012,Marcuzzi2014b}. Moreover, this FRG approach can provide quantitative predictions for additional relevant non-equilibrium universal quantities such as the fluctuation-dissipation ratio and the effective temperatures in the aging regime~\cite{Calabrese2002,Calabrese2002b,Calabrese2004,Calabrese2005,Cugliandolo2011}.

Finally, the approach discussed here constitutes a first step towards the exploration of universality in the dynamics of isolated quantum many-body systems after a parameter quench, a current topic of considerable theoretical and experimental interest~\cite{PolkovnikovRMP,Dalessio2015,Eisert2015b,Essler2016,Langen2016,Calabrese2016,Caux2016}. 

\begin{acknowledgments}
We thank A.~Codello, N.~Defenu, J.~Krug, S.-C.~Park, R.~Percacci, P.~Politi, D.~Roscher, M.~Schir\'o and A.~Trombettoni for useful discussions. A.~C. acknowledges the kind hospitality of the Technische Universit{\"a}t Dresden and the Institut f\"{u}r Theoretische Physik at the University of Cologne, where this work was partially done.
J.~M. acknowledges support from the Alexander Von Humboldt foundation. S. D. acknowledges funding by the German Research Foundation (DFG) through the Institutional Strategy of the University of Cologne within the German Excellence Initiative (ZUK 81), and by the European Research Council via ERC Grant Agreement n. 647434 (DOQS).
\end{acknowledgments}
%


\appendix


%
\section{Derivation of the FRG equation}
\label{app:wetterich}
In this Appendix we briefly review the derivation of Eq.~\eqref{eq:wetterich} for the response functional~\cite{Berges2002}.
Let us consider the action $S[\Psi]$, where $\Psi^\text{t} = (\varphi,\widetilde{\varphi})$, with $\varphi$ the order parameter and $\widetilde{\varphi}$ the response field. We define a modified action $S_k[\Psi] = S[\Psi] + \Delta S_k[\Psi]$ where $\Delta S_k[\Psi] =\frac{1}{2}\int_{t,\rr} \Psi^\text{t} \sigma \Psi R_k$, where $\sigma =\left(\begin{smallmatrix} 0 & 1 \\1 & 0\end{smallmatrix}\right)$ and $R_k$ is a function which implements the infrared cutoff. Then we define the generating function $W_k[J]$ as
\begin{equation}
\label{eq:W-definition}
W_k[J] = \log \left[\int \mathrm{D}\Psi\, \ee^{-S_k[\Psi] + \int_{t,\rr}\Psi^\text{t} J}\right],
\end{equation}
where $\mathrm{D}\Psi$ denotes functional integration over both the fields $\varphi$ and $\widetilde{\varphi}$, while $J^\text{t}=(j,\widetilde{j})$ is an external field. 
Defining the expectation value $\langle \Psi \rangle$, where the average is taken with respect to the action $-S_k[\Psi] + \int_{t,\rr}\Psi^\text{t} J$, it is straightforward to check that the following properties follows from Eq.~\eqref{eq:W-definition}~\cite{Berges2002}: 
\begin{equation}
\label{eq:W-properties}
\langle \Psi \rangle = \frac{\delta W_k}{\delta J^\text{t}}, \qquad \langle \Psi \Psi^\text{t} \rangle - \langle \Psi \rangle \langle \Psi^\text{t} \rangle =\frac{\delta^2 W_k }{\delta J^\text{t}\delta J} =\frac{\delta \langle \Psi^\text{t} \rangle}{\delta J^\text{t}}.
\end{equation}
The effective action $\Gamma_k[\Phi]$ is defined as
\begin{equation}
\label{eq:Gamma-definition}
\Gamma_k[\Phi] = -W_k[J] + \int_{t,\rr} J^\text{t} \Phi - \Delta S_k[\Phi],
\end{equation}
where $J$ is fixed by the condition
\begin{equation}
\Phi = \frac{\delta W_k}{\delta J^\text{t}}.
\end{equation}
By comparing the previous equation with the first one in Eq.~\eqref{eq:W-properties}, it follows that $\Phi =\langle \Psi \rangle$: accordingly, by using Eq.~\eqref{eq:W-properties}, the following relationships can be derived~\cite{Berges2002}: 
\begin{equation}
\label{eq:Gamma-properties}
\frac{\delta \Gamma_k}{\delta \Phi^\text{t}} = J - \sigma R_k\Phi, \qquad \frac{\delta^2 \Gamma_k}{\delta \Phi^\text{t} \delta\Phi} + \sigma R_k = \frac{\delta J^\text{t}}{\delta\Phi ^\text{t}} = \left[\frac{\delta^2 W_k }{\delta J^\text{t}\delta J}\right]^{-1}.
\end{equation}
The definition of $\Gamma_k[\Phi]$ in Eq.~\eqref{eq:Gamma-definition} is such that~\cite{Berges2002} $\Gamma_{k=\Lambda}[\Phi] \approx S[\Phi]$, i.e., when $k$ is equal to the ultraviolet cutoff $\Lambda$ of the theory, the effective action $\Gamma_k$ reduces to the ``microscopic'' action $S[\Phi]$ evaluated on the expectation value $\Phi$. This can also be easily seen by taking a Gaussian microscopic action $S[\Psi]$: in this case a simple calculation shows that $\Gamma_k[\Phi] = S[\Phi]$. 
We can now derive the FRG equation by taking the total derivative of the effective action with respect to $k$:
\begin{align}
\frac{\dd \Gamma_k}{\dd k} & = -\frac{\partial W_k}{\partial k} - \int_{t,\rr}\left(\frac{\delta W}{\delta J}-\Phi^\text{t}\right)\frac{\dd J}{\dd k}  - \frac{1}{2}\int_{t,\rr} \Phi^\text{t} \sigma \frac{\dd R_k}{\dd k} \Phi \nonumber \\
& = \frac{1}{2} \int_{t,\rr} \langle \Psi^\text{t} \sigma \frac{\dd R_k}{\dd k} \Psi\rangle - \frac{1}{2}\int_{t,\rr} \Phi^\text{t} \sigma \frac{\dd R_k}{\dd k} \Phi \nonumber \\
& = \frac{1}{2} \int_{t,\rr} \text{tr}\left[\langle \Psi\Psi^\text{t}\rangle \sigma \frac{\dd R_k}{\dd k} \right] - \frac{1}{2}\int_{t,\rr} \text{tr}\left[ \Phi\Phi^\text{t} \sigma\frac{\dd R_k}{\dd k} \right] \nonumber \\
& =\frac{1}{2} \int_{t,\rr} \text{tr}\left[\left(\frac{\delta^2 \Gamma_k}{\delta \Phi^\text{t} \delta\Phi}+\sigma R_k\right)^{-1}\sigma \frac{\dd R_k}{\dd k} \right], \label{eq:wetterich-derivation}
\end{align}
where we repeatedly used Eqs.~\eqref{eq:W-properties} and~\eqref{eq:Gamma-properties} and we expressed the scalar products $\Psi^\text{t} \sigma \Psi$ and $\Phi^\text{t} \sigma \Phi$ as traces over the internal degrees of freedom.
Equation~\eqref{eq:wetterich-derivation} is the FRG equation which describes the flow of the effective action $\Gamma_k$ upon varying the infrared cutoff $k$.

\section{Derivation of Gaussian Green's functions from $\Gamma_0$}
\label{app:dyson-initial}
In this Appendix we show how the boundary action $\Gamma_0$ in Eq.~\eqref{eq:boundary-action} contributes to the matrix $G_0$ defined in Eq.~\eqref{eq:G0-definition}. Let us consider the quadratic part of the effective action (we consider $Z_0 = 1$ and $h_0 = 0$ for the sake of simplicity) expressed in momentum space:
\begin{align}
\label{eq:action-quadratic}
\Gamma	 & = \int_\qq \left( -\frac{1}{2\tau_0}\res_0^2 + \res_0\phi_0 \right) \nonumber \\
				 & \qquad  + \int_{t,\qq} \vartheta(t-t_0)\,\res \left( \dot{\phi} + \omega_q\phi  - D\res\right), 
\end{align}
where $\omega_q = q^2 + \tau$ is the dispersion law, $\int_\qq \equiv \int \dd^d q/(2\pi)^d$ and $\phi \equiv \phi(t,\qq), \res \equiv \res(t,\qq)$. 
By taking its second variation $\Gamma^{(2)}(q,t,t')$ as defined in Eq.~\eqref{eq:Gamma-second-variation}, one finds
\begin{equation}
\label{eq:second-variation-time}
\Gamma^{(2)}(q,t,t') = \left[- V_0\,\delta(t-t_0) + \hat{B}_q(t)\right]\delta(t-t'),
\end{equation}
where the matrices $V_0$  and $\hat{B}_q(t)$ are defined as
\begin{equation}
\label{eq:matrices}
V_0 = \begin{pmatrix}
0 & -1 \\
-1& \tau_0^{-1}
\end{pmatrix}, \qquad 
\hat{B}_q(t) = \begin{pmatrix}
0 & -\partial_t +\omega_q \\
\partial_t+\omega_q & -2 D
\end{pmatrix}.
\end{equation}
The matrix $V_0$ is obtained from the boundary action, and consequently it appears in Eq.~\eqref{eq:second-variation-time} multiplied by a delta function localized at $t=t_0$, while the term proportional to $\hat{B}_q(t)$ is, instead, related to the bulk action.
The matrix $G_0(t,t')$ of the correlation functions, defined in Eq.~\eqref{eq:G0-definition}, is given by $G_0(q,t,t') = [\Gamma^{(2)}]^{-1}(q,t,t')$, where the inverse is taken with respect to the internal matrix structure, the times $t,t'$, and the momentum $q$. However, since the matrix is diagonal in $q$, the inversion with respect to the dependence on momenta is trivial. Making use of the definition of $G(q,t,t')$ in Eq.~\eqref{eq:G-definition}, multiplying both sides by $G_\text{eq}(q,t,t') \equiv [\hat{B}_q(t,t')]^{-1}$ and integrating over the intermediate times, one finds
\begin{equation}
\label{eq:dyson-time}
G_0(t,t') = G_\text{eq}(t,t') + G_\text{eq}(t,t_0)V_0G_0(t_0,t').
\end{equation} 
The explicit form of $G_\text{eq}(q,t,t')$ can be calculated by inverting the Fourier transform of $\hat{B}_q(t,t')$ and anti-transforming in real time:
\begin{equation}
\begin{split}
\label{eq:equilibrium-greens}
&G_\text{eq}(q,t,t') = \int \frac{\dd\omega}{2\pi}\, \ee^{-i\omega (t-t')} [B_q(\omega)]^{-1} = \\
&=\begin{pmatrix}
D\omega_q^{-1}\ee^{-\omega_q|t-t'|} & \vartheta(t-t')\ee^{-\omega_q(t-t')} \\
\vartheta(t'-t)\ee^{-\omega_q(t'-t)} & 0
\end{pmatrix}.
\end{split}
\end{equation}
Notice that $B_q(\omega)$ is diagonal in the frequency $\omega$, since $\hat{B}_q(t,t')$ depends only on the difference of times $t-t'$. In fact, $G_\text{eq}(t,t')$ is a time-translational invariant function which corresponds to the correlation matrix of the model at thermal equilibrium. 
Equation~\eqref{eq:dyson-time} can be solved by iteration, and it yields the formal solution:
\begin{multline}
\label{eq:dyson-formal-solution}
G_0(q,t,t')  = G_\text{eq}(q,t,t') + G_\text{eq}(q,t,t_0) V_0 \\
\times \sum_{n=0}^{+\infty}\left[ G_\text{eq}(q,t_0,t_0)V_0 \right]^n G_\text{eq}(q,t_0,t').
\end{multline}
Recalling that within the response functional formalism adopted here we set $\vartheta(0) = 0$ in order to ensure causality~\cite{Tauberbook2014}, we get 
\begin{equation}
\begin{split}
\label{eq:summation}
&V_0\sum_{n=0}^{+\infty}\left[ G_\text{eq}(q,t_0,t_0)V_0 \right]^n = \\
& = V_0 [1-G_\text{eq}(q,t_0,t_0)V_0]^{-1}  = 
\begin{pmatrix}
0  & -1 \\
-1 & D\omega_q^{-1} + \tau_0^{-1} 
\end{pmatrix}.
\end{split}
\end{equation}
Combining Eqs.~\eqref{eq:dyson-formal-solution}, \eqref{eq:equilibrium-greens} and~\eqref{eq:summation}, one finds the same Gaussian Green's functions as those in Eqs.~\eqref{eq:GK-gaussian} and~\eqref{eq:GR-gaussian}, with $\Omega = 1$ and $T$ replaced by $D$.

\section{Integral equation for $G$}
\label{app:Dyson}
In this Appendix we derive and discuss Eq.~\eqref{eq:Dyson-maintext} for the matrix $G$ defined in Eq.~\eqref{eq:G-definition}.
The former can be obtained by multiplying both sides of the latter by $G_0^{-1}-V$ defined in Eq.~\eqref{eq:G0-V-definition} and by integrating over intermediate coordinates, which yields
\begin{equation}
\label{eq:G-Gamma}
\int_y \left[ G^{-1}_0(x,y) - V(x,y) \right] G(y,x') = \delta(x-x'), 
\end{equation}
where the delta function on the r.h.s.~of Eq.~\eqref{eq:G-Gamma} appears as a consequence of Eq.~\eqref{eq:G-G0-V}. Accordingly, by multiplying both sides of Eq.~\eqref{eq:G-Gamma} by $G_0$ and integrating over the intermediate coordinates, and by using Eq.~\eqref{eq:self-energy}, one finds the integral equation for $G$
\begin{align}
\label{eq:Dyson}
G(x,x') 	 = G_0(x,x') + \int_y\, G_0(x,y)\, V(y) G(y,x').
\end{align}
This equation can be formally solved by iteration, and the solution can be expressed as the infinite series 
\begin{equation}
\label{eq:Dyson-series}
G(x,x') = G_0(x,x') + \sum_{n=1}^{+\infty} G_{n}(x,x'),
\end{equation}
where $G_{n}$ are convolutions given by
\begin{align}
\label{eq:Gn-definition}
G_{n}(x,x')
& \equiv \int_{y_1\dots y_n} G_0(x,y_1)V(y_1)G_0(y_1,y_2) \nonumber \\
& \qquad \qquad  \times\dots V(y_n)G_0(y_n,x'). 
\end{align}
The formal solution Eq.~\eqref{eq:Dyson-series} can be inserted into Eq.~\eqref{eq:wetterich}, providing a convenient expression for the FRG equation, which now reads (as in the main text, the dependence of $\Gamma$ and $R$ on $k$ is understood):
\begin{equation}
\label{eq:wetterich-series}
\frac{\dd \Gamma}{\dd k} = \sum_{n=0}^{+\infty} \Delta \Gamma_n, 
\end{equation}
where
\begin{equation}
\label{eq:DeltaGamma-0}
\Delta\Gamma_0 \equiv \frac{1}{2} \int_x  \text{tr}\left[ G_0(x,x)\frac{\dd R}{\dd k}\sigma\right],
\end{equation}
and
\begin{equation}
\label{eq:DeltaGamma-n}
\Delta \Gamma_n = \frac{1}{2} \int_x  \text{tr}\left[ G_{n}(x,x)\frac{\dd R}{\dd k}\sigma\right] \;\; \text{for} \;\; n \ge 1.
\end{equation}
A straightforward calculation shows that $\Delta\Gamma_0 \propto \vartheta(0)$ and therefore this term vanishes, since we assumed from the outset $\vartheta(0) = 0$ in order to ensure causality~\cite{Tauberbook2014}. As a result, the sum over $n$ in Eq.~\eqref{eq:wetterich-series} actually starts from $n=1$, as in Eq.~\eqref{eq:wetterich-series-maintext}.

The FRG equation in the form of Eq.~\eqref{eq:wetterich-series} can be used to study the case of systems with broken time-translational symmetry, since each $\Delta \Gamma_n$ can now be calculated independently of the presence of such a symmetry. However, we emphasize that in general it is not possible to sum the series on the r.h.s.~of Eq.~\eqref{eq:wetterich-series} in a closed form, because the convolutions in $\Delta\Gamma_n$ are generically rather complicated non-local functions of the fields $\phi$ and $\res$.

If, instead, time-translational invariance is not broken, e.g., when one takes the limit $t_0 \to -\infty$ in the action~\eqref{eq:effective-action}, the matrix $G$
determined from Eq.~\eqref{eq:Dyson-series} is identical to the one obtained by the direct inversion of Eq.~\eqref{eq:G-definition}.
In order to show this, let us assume that one is interested only in the renormalization of the potential $\mathcal{U}$, disregarding those of $K, D$ and $Z$. Then one makes use of the so-called local-potential approximation~\cite{Berges2002, Delamotte2007,Canet2005}, in which the field-dependent function $V(\rr,t)$ introduced in Eq.~\eqref{eq:V-definition} is evaluated on configurations of the fields $\phi$ and $\res$ which are constant in space and time, such that $V(\rr,t)$ is actually independent of $\rr$ and $t$.
As a consequence of time-translational invariance, $G$ depends only on the difference of its arguments, i.e., $G_0(x,x') = G_0(x-x')$ and $G(x,x') = G(x-x')$. Then, after taking the Fourier transform with respect to the relative coordinates $\rr-\rr'$ and $t-t'$, the convolutions in Eq.~\eqref{eq:Dyson-series} become products of the $G_0(\kk,\omega)$, which are functions of the momentum $\kk$ and of the frequency $\omega$. Accordingly, this equation becomes
\begin{align}
\label{eq:G-resummed}
G(\kk,\omega) & = \sum_{n=0}^{+\infty} G_0(\kk,\omega) \left[ V G_0(\kk,\omega) \right]^n \nonumber \\
						& = [G^{-1}_0(\kk,\omega) - V]^{-1}.
\end{align}
This expression can thus be used in Eq.~\eqref{eq:wetterich}, which then acquires a closed form.
Notice that Eq.~\eqref{eq:G-resummed} could have been obtained directly  by simply taking the Fourier transform of Eq.~\eqref{eq:G-definition} with the definition~\eqref{eq:G0-V-definition}.   \\

\section{Calculation of $\Delta\Gamma_1$ and $\Delta \Gamma_2$}
\label{app:evaluation-DGamma}
In this Appendix, we detail the calculations which lead to Eqs.~\eqref{eq:dGamma-1} and~\eqref{eq:dGamma-2}.
Starting from  Eq.~\eqref{eq:DeltaGamma-n}, we find:
\begin{align}
\label{eq:DeltaGamma1-computation}
&\Delta \Gamma_1	= \nonumber \\
& \, = \frac{1}{2} \int_{t,t',\rr,\rr'} \!\!\!\!\!\!\!  \text{tr}\left[G_0(\rr-\rr',t,t')\,V(t',\rr')G_0(\rr'-\rr,t',t)\sigma\frac{\dd R}{\dd k}\right] \nonumber \\
& \, = \frac{1}{2} \int_{t,t',\qq,\rr'} \!\!\!\!\!\!\! \text{tr}\left[G_0(q,t,t')\,V(t',\rr')G_0(q,t',t)\sigma\frac{\dd R}{\dd k}(q)\right] \nonumber \\
& \, = k^{d+1} \frac{a_d}{d} \int_{t,t',\rr'} \text{tr}\left[G_0(k,t,t')\,V(t',\rr')G_0(k,t',t)\sigma\right],
\end{align}
where $a_d = 2/[\Gamma(d/2) (4\pi)^{d/2}]$, with $d$ the spatial dimensionality and $\Gamma(x)$ the gamma function.
In the second equality of Eq.~\eqref{eq:DeltaGamma1-computation} one expresses $G_0(\rr,t,t')$ in terms of its Fourier transforms $G_0(q,t,t')$ and then calculates the integral over the spatial coordinates $\rr$. 
In the third equality, instead, after performing the integration over angular variables (which generates the factor $a_d$), the integral over momenta $q$ becomes trivial since the function (see Eq.~\eqref{eq:Litim})
\begin{equation}
\label{eq:R-derivative}
\frac{\dd R (q)}{\dd k} = 2k\,\vartheta(k^2 - q^2)
\end{equation}
restricts the integration domain to $0\leq q \leq k$, within which $G_0(q,t,t')$ is constant and equal to $G_0(k,t,t')$ as a consequence of the modified dispersion relation in Eq.~\eqref{eq:modified-dispersion}. Similarly, $\omega_{k,q}$ is replaced by $\omega_{k,q\leq k} = \omega_{k,k} = \omega_{q=k}$ (see Eq.~\eqref{eq:modified-dispersion} and after Eq.~\eqref{eq:GK-gaussian}). Note that, since $K$ is not renormalized within this approximation, it does not contribute to Eq.~\eqref{eq:R-derivative} and, for simplicity, we set $K=1$.
Finally, by using the definitions~\eqref{eq:G0-definition} and~\eqref{eq:self-energy-quartic}, one evaluates the trace in Eq.~\eqref{eq:DeltaGamma1-computation}, finding
\begin{align}
\Delta \Gamma_1	
& = -2k^{d+1} \frac{a_d}{d} g\, \int_{\rr'}\int_{t_0}^{+\infty} \dd t'\, \res(\rr',t')\phi(\rr',t') \nonumber \\
& \qquad \qquad \times \int_{t_0}^{+\infty} \dd t\, G_R(k,t',t)G_C(k,t',t) \nonumber \\
& = -k^{d+1}\frac{a_d}{d}\frac{g D}{\omega_k^2} \int_{\rr}\int_{t_0}^{+\infty} \dd t'\, \res(t',\rr)\phi(t',\rr) \nonumber \\
& \qquad \qquad \times \left[ 1 - f_\tau(t'-t_0)\right], \label{eq:Delta-Gamma-1-calculation}
\end{align}
where in the last equality the integral over time $t$ was calculated. The function $f_\tau(t)$ is defined in Eq.~\eqref{eq:f-definition} and corresponds to the time-dependent part of the result of the integration over $t$. Note that the terms proportional to $\phi^2$ contained in $V$, do not appear in the final result (as required by causality \cite{Kamenevbook2011, Tauberbook2014}) since they would be multiplied by a factor $\vartheta(t-t')\vartheta(t'-t) = 0$.  The last equality of Eq.~\eqref{eq:Delta-Gamma-1-calculation} is nothing but Eq.~\eqref{eq:dGamma-1} of the main text.

The calculation of $\Delta \Gamma_2$ is lengthier, but it proceeds as discussed above for $\Delta \Gamma_1$. From the definition in Eq.~\eqref{eq:DeltaGamma-n} one has
\begin{widetext}
\begin{align}
\label{eq:DeltaGamma2-computation-1}
\Delta \Gamma_2	
& = 	\frac{1}{2}\int_{t,t',t'',\rr,\rr',\rr''} \text{tr}\left[G_0(\rr-\rr',t,t')V(t',\rr')G_0(\rr'-\rr'',t',t'')V(t'',\rr'')G_0(\rr''-\rr,t'',t)\frac{\dd R}{\dd k}\right] \nonumber \\
& = \frac{1}{2}\int_{t,t',t'',\rr',\rr'',\qq,\qq'} \ee^{i\rr' \cdot(\qq'-\qq)} \, \text{tr}\left[G_0(q,t,t')V(t',\rr'+\rr'')G_0(q',t',t'')V(t'',\rr'')G_0(q,t'',t)\frac{\dd R(q)}{\dd k}\right] \nonumber \\					& 	\approx \frac{1}{2}\int_{t,t',t'',\rr',\rr'',\qq,\qq'} \ee^{i\rr' \cdot(\qq'-\qq)} \text{tr}\left[G_0(q,t,t')V(t',\rr'')G_0(q',t',t'')V(t'',\rr'')G_0(q,t'',t)\frac{\dd R(q)}{\dd k}\right]  \nonumber \\	
& = k^{d+1}\frac{a_d}{d}\int_{t,t',t'',\rr''} \text{tr}\left[G_0(k,t,t')V(t',\rr'')G_0(k,t',t'')V(t'',\rr'')G_0(k,t'',t)\sigma\right],
\end{align}
\end{widetext}  
where in the second equality we expressed the various $G_0(\rr,t,t')$ (see Eq.~\eqref{eq:G0-definition}) in terms of their Fourier transforms, we made the change of variables $\rr' \to \rr' + \rr''$ and integrated over the spatial coordinate $\rr$. In the third step we expanded $V(t',\rr'+\rr'') \approx V(t',\rr'')$ in order to retain only local combinations of fields, while in the last step we integrated over $\rr'$ and calculated the trivial integrals over $\qq$ and $\qq'$.  
After determining the trace on the basis of the definitions~\eqref{eq:G0-definition} and~\eqref{eq:self-energy-quartic}, and by noticing that the prefactor of the term $\propto \phi^4$ vanishes (as required by causality \cite{Kamenevbook2011, Tauberbook2014}) as it contains the factor $\vartheta(t-t')\vartheta(t'-t'')\vartheta(t''-t) = 0$, Eq.~\eqref{eq:DeltaGamma2-computation-1} becomes
\begin{widetext}
\begin{align}
\label{eq:DeltaGamma2-computation-2}
\Delta \Gamma_2 
& = k^{d+1}\frac{a_d}{d}g^2\int_{\rr''}\int_{t_0}^{+\infty} \dd t\, \dd t' \dd t''\, \bigg\{ 2\, \res(t')\phi(t')\res(t'')\phi(t'') \, G_{0C}(k,t',t'')G_{0C}(k,t'',t)G_{0R}(k,t',t) \nonumber \\ 
& \qquad  + \res(t')\phi(t')\phi^2(t'') \bigg[ G_{0C}(k,t',t'')G_{0R}(k,t,t'')G_{0R}(k,t',t) \nonumber \\
& \qquad + G_{0C}(k,t',t'')G_{0R}(k,t,t'')G_{0R}(k,t',t) + G_{0C}(k,t,'t)G_{0R}(k,t',t'')G_{0R}(k,t'',t) \bigg]\bigg\} \nonumber \\
& \simeq  k^{d+1}\frac{a_d}{d}g^2\int_{\rr''}\int_{t_0}^{+\infty} \dd t'\, \left[ 2\res^2(t')\phi^2(t') F_D(t')  + \res(t')\phi^3(t') F_g(t')\right],
\end{align}
\end{widetext}
where we omitted the dependence on $\rr''$ of the fields for the sake of clarity. In the last step of Eq.~\eqref{eq:DeltaGamma2-computation-2}, we expanded the fields for $t'\simeq t''$ as $\phi(t'') \simeq \phi(t')$ and $\res(t'') \simeq \res(t')$ in order to retain only combinations of the fields local in time, and we introduced the functions
\begin{equation}
\label{eq:FD-definition}
F_D(t') = \int_{t_0}^{+\infty} \dd t\,\dd t'' G_{0C}(k,t,t')G_{0C}(k,t',t'')G_{0R}(k,t'',t),
\end{equation}
and
\begin{align}
\label{eq:Fg-definition}
F_g(t') 
& = \int_{t_0}^{+\infty}\dd t\,\dd t'' \bigg[ G_{0C}(k,t',t'')G_{0R}(k,t,t'')G_{0R}(k,t',t) \nonumber \\
&  \quad\quad\quad + G_{0C}(k,t,t'')G_{0R}(k,t',t'')G_{0R}(k,t',t) \nonumber \\
&  \quad\quad\quad + G_{0C}(k,t,t')G_{0R}(k,t',t'')G_{0R}(k,t'',t)\bigg].
\end{align}
The functions $F_D$ and $F_g$ can be easily evaluated using Eqs.~\eqref{eq:GK-gaussian} and~\eqref{eq:GR-gaussian}, and they render
\begin{align}
F_D(t) & = \frac{1}{4}\frac{D^2}{\omega^4_k}\left[3 - f_D(t-t_0)\right], \label{eq:FD-expression}\\
F_g(t) & = \frac{D}{\omega_k^3}\left[1 - f_g(t-t_0)\right], \label{eq:Fg-expression}
\end{align}
with 
\begin{align}
\label{eq:fD}
f_D(t) & =  \left\{ 2 + 2\omega_k t - 2(1+\omega_k t)^2\left( \frac{\omega_k}{D\tau_0} - 1 \right) \right. \nonumber \\
& \qquad \qquad - \left. \left( \frac{\omega_k}{D\tau_0} - 1 \right)^2\right\} \ee^{-2\omega_k t},\\
f_g(t) & = \left[1+2\omega_k t - 2(\omega_k t)^2\left(\frac{\omega_k}{D\tau_0}-1\right)\right]\ee^{-2\omega_k t}. \label{eq:fg}
\end{align}
Finally, substituting Eqs.~\eqref{eq:FD-expression} and~\eqref{eq:Fg-expression} into Eq.~\eqref{eq:DeltaGamma2-computation-2}, we find Eq.~\eqref{eq:dGamma-2} of the main text.

\section{Flow equations in the ordered phase}
\label{app:beta_ordered}
In this Appendix, we will detail the derivation of the flow equations for the potential expanded around a finite homogeneous value $\phi=\phi_m$ and $\res=0$. For the sake of clarity, we consider the potential $\mathcal{U}$ in Eq.~\eqref{eq:potential-ordered} with $\lambda = 0$. The generalization to the case $\lambda \neq 0$ is straightforward and proceeds as in the equilibrium case (see, for instance, Refs.~\onlinecite{Berges2002} and~\onlinecite{Sieberer2014}).

First of all, since the factor $K$ is renormalized within the ansatz discussed here, the derivative with respect to $k$ of the regulator $R(q)$ defined in Eq.~\eqref{eq:Litim}, has also to account for the renormalization factor $K$ on $k$, as
\begin{equation}
\label{eq:regulator-derivative}
\frac{\dd R (q)}{\dd k} = \frac{K}{k}\vartheta(k^2-q^2)\left[2k^2 - \eta_K\left(k^2-q^2\right)\right],
\end{equation}
where we made use of the definition of $\eta_K$ in Eq.~\eqref{eq:anomalous-dimensions}, see also Ref.~\onlinecite{Sieberer2014}. In fact, since the factor $K$ depends on $k$ within this approximation, the derivative with respect to $k$ of Eq.~\eqref{eq:Litim} produces a contribution proportional to $\eta_K$.

Then, by taking the second variation of the effective action $\Gamma$ in Eq.~\eqref{eq:effective-action} (see Eq.~\eqref{eq:Gamma-second-variation}), we  cast the equation for the function $G$ defined in Eq.~\eqref{eq:G-definition} into the same form as Eq.~\eqref{eq:Dyson}, with the field-dependent function $V$ defined as (we assume $t_0 = 0$ for simplicity): 
\begin{equation}
\label{eq:self-energy-ordered}
V(x) = -g\,  \vartheta(t)
\begin{pmatrix}
\widetilde{\rho}(x)		& \rho(x)-\rho_m \\
\rho(x)-\rho_m 	&	 0
\end{pmatrix},
\end{equation}
where we define
\begin{equation}
\label{eq:invariants}
\rho\equiv\frac{\phi^2}{2}, \qquad \widetilde{\rho}=\res\phi, \qquad \rho_m \equiv \frac{\phi_m^2}{2}, 
\end{equation}
while $G_0$ is defined according to Eq.~\eqref{eq:G0-definition}, but with the post-quench parameter $r$ replaced by
\begin{equation}
\label{eq:effective-mass}
m=\frac{2}{3}\rho_m g.
\end{equation}
The use of the $\mathbb{Z}_2$ invariants $\rho$ and $\widetilde{\rho}$ is customary in the context of FRG~\cite{Sieberer2014} and it helps in simplifying the notation in what follows. 
The form of $V(x)$ in Eq.~\eqref{eq:self-energy-ordered} allows us to express the r.h.s.~of the FRG equation~\eqref{eq:wetterich-series-maintext} as a power series of $\rho-\rho_m$, in the  spirit of Eq.~\eqref{eq:wetterich-series}: this provides, together with the vertex expansion \eqref{eq:vertexp-maintext}, a way to unambiguously identify the renormalization of the terms appearing in the potential $\mathcal{U}$ in Eq.~\eqref{eq:potential-ordered}. In fact, $\rho_m$ and the couplings $g$ and $\lambda$ are identified as~\cite{Delamotte2007}:
\begin{equation}
\label{eq:projections-rules-1}
\frac{\dd \mathcal{U}}{\dd \rho}\biggr|_{\rho = \rho_m} \!\!\!\!\! = 0, \quad 
\frac{g}{3} = \frac{\dd^2 \mathcal{U}}{\dd \rho^2}\biggr|_{\rho=\rho_m}, \quad 
 \frac{\lambda}{15} = \frac{\dd^3 \mathcal{U}}{\dd \rho^3}\biggr|_{\rho=\rho_m},
\end{equation}
where the first condition actually defines $\rho_m$ as the minimum of the potential. In terms of the effective action $\Gamma$, Eqs.~\eqref{eq:projections-rules-1} become~\cite{Canet2007,Sieberer2014}
\begin{equation}
\label{eq:projections-rules-2}
\frac{\delta \Gamma}{\delta \widetilde{\rho}}\biggr|_{\substack{\widetilde{\rho}=0\\ \rho=\rho_m}} \!\!\!\!\!= 0, \quad 
\frac{g}{3} = \frac{\delta^2 \Gamma }{\delta\widetilde{\rho}\, \delta\rho}\biggr|_{\substack{\widetilde{\rho}=0\\ \rho=\rho_m}}, \quad  
\frac{\lambda}{15} = \frac{\delta^3 \Gamma}{\delta\widetilde{\rho}\, \delta\rho^2}\biggr|_{\substack{\widetilde{\rho}=0\\ \rho=\rho_m}}.
\end{equation}
By taking a total derivative with respect to $k$ of each equality in Eqs.~\eqref{eq:projections-rules-2}, one finds
\begin{align}
& \frac{\delta }{\delta \widetilde{\rho}} \frac{\partial\Gamma}{\partial k}\biggr|_{\substack{\widetilde{\rho}=0\\ \rho=\rho_m}} + \frac{\delta^2 \Gamma}{\delta \widetilde{\rho}\,\delta \rho}\biggr|_{\substack{\widetilde{\rho}=0\\ \rho=\rho_m}} \frac{\dd \rho_m}{\dd k}=0, \label{eq:rho0-flow-general}\\ 
& \frac{1}{3} \frac{\dd g}{\dd k} = \frac{\delta^2 }{\delta \widetilde{\rho}\,\delta \rho} \frac{\partial\Gamma}{\partial k}\biggr|_{\substack{\widetilde{\rho}=0\\ \rho=\rho_m}} + \frac{\delta^3 \Gamma}{\delta \widetilde{\rho}\,\delta^2 \rho}\biggr|_{\substack{\widetilde{\rho}=0\\ \rho=\rho_m}} \frac{\dd \rho_m}{\dd k},\label{eq:g-flow-general}\\
& \frac{1}{15} \frac{\dd \lambda}{\dd k} = \frac{\delta^3 }{\delta \widetilde{\rho}\,\delta \rho^2} \frac{\partial\Gamma}{\partial k}\biggr|_{\substack{\widetilde{\rho}=0\\ \rho=\rho_m}} + \frac{\delta^4 \Gamma}{\delta \widetilde{\rho}\,\delta^3 \rho}\biggr|_{\substack{\widetilde{\rho}=0\\ \rho=\rho_m}} \frac{\dd \rho_m}{\dd k},\label{eq:lambda-flow-general}
\end{align}
which, after replacing $\partial \Gamma /\partial k$ with the FRG equation~\eqref{eq:wetterich}, render the flow equations for $\rho_m$, $g$, and $\lambda$. For the case of the potential $\mathcal{U}$ in Eq.~\eqref{eq:potential-ordered} with $\lambda = 0$, by using Eq.~\eqref{eq:projections-rules-2}, the set of flow equations~\eqref{eq:rho0-flow-general} and~\eqref{eq:g-flow-general} simplifies as
\begin{align}
\frac{\dd \rho_m}{\dd k} &= - \frac{3}{g}\frac{\delta }{\delta \widetilde{\rho}} \frac{\partial\Gamma}{\partial k}\biggr|_{\substack{\widetilde{\rho}=0\\ \rho=\rho_m}} = - \frac{3}{g}\frac{\delta \Delta \Gamma_1}{\delta \widetilde{\rho}}  \biggr|_{\substack{\widetilde{\rho}=0\\ \rho=\rho_m}}, \label{eq:rho0-flow-ordered}\\
\frac{1}{3} \frac{\dd g}{\dd k} & = \frac{\delta^2 }{\delta \widetilde{\rho}\,\delta \rho} \frac{\partial\Gamma}{\partial k}\biggr|_{\substack{\widetilde{\rho}=0\\ \rho=\rho_m}} = \frac{\delta^2\Delta\Gamma_2 }{\delta \widetilde{\rho}\,\delta \rho} \biggr|_{\substack{\widetilde{\rho}=0\\ \rho=\rho_m}}\label{eq:g-flow-ordered},
\end{align}
where we used Eq.~\eqref{eq:wetterich-series} with $\Delta \Gamma_1$ and $\Delta \Gamma_2$ defined as in Eqs.~\eqref{eq:DeltaGamma-n} and~\eqref{eq:Gn-definition} in terms of the $V(x)$ in Eq.~\eqref{eq:self-energy-ordered}.  
The explicit form of the flow equations comes from a calculation analogous to the one discussed in Sec.~\ref{sec:truncation-symmetric} and in App.~\ref{app:evaluation-DGamma} (see Eqs.~\eqref{eq:dGamma-1} and~\eqref{eq:dGamma-2}). In particular, the flow of $m$, defined in Eq.~\eqref{eq:effective-mass}, takes contributions from both the flow equations for $\rho_m$ and $g$. Similarly, the renormalization of $Z_0$ is determined by the contribution localized at $t=0$ of the coefficient of the quadratic term $\res\phi$ in the effective action~\eqref{eq:effective-action} equipped with the potential~\eqref{eq:potential-ordered}.

\section{Anomalous dimensions}
\label{app:anomalous-dimensions}
In this Appendix we discuss the derivation of the renormalization of $K$, $Z$, and $D$ resulting from the potential $\mathcal{U}$ in Eq.~\eqref{eq:potential-ordered} and from the effective action $\Gamma$ in Eq.~\eqref{eq:effective-action}.

\subsection{Renormalization of $D$}
\label{app:anomalousD}
The strength $D$ of the Markovian noise can be unambiguously defined from the effective action $\Gamma$ in Eq.~\eqref{eq:effective-action} as~\cite{Sieberer2014}
\begin{equation}
D = -\rho_m\, \frac{\delta^2 \Gamma}{\delta^2\widetilde{\rho}}\biggr|_{\substack{\widetilde{\rho}=0\\ \rho=\rho_m}},
\end{equation}
where $\rho$ and $\widetilde{\rho}$ are defined in Eq.~\eqref{eq:invariants}.
By differentiating the previous equation with respect to $k$, we find
\begin{equation}
\label{eq:D-flow-general}
\frac{\dd D}{\dd k} = -\left( \frac{\delta^2 \Gamma}{\delta^2\widetilde{\rho}} + \frac{\delta^3 \Gamma}{\delta^2\widetilde{\rho}\delta\rho} \right)\biggr|_{\substack{\widetilde{\rho}=0\\ \rho=\rho_m}} \frac{\dd \rho_m}{\dd k} - \rho_m \frac{\delta^2 }{\delta^2\widetilde{\rho}} \frac{\partial \Gamma}{\partial k}\biggr|_{\substack{\widetilde{\rho}=0\\ \rho=\rho_m}}.
\end{equation}
For the effective action $\Gamma$ in Eq.~\eqref{eq:effective-action} with the potential $\mathcal{U}$ in Eq.~\eqref{eq:potential-ordered}, the terms in brackets in Eq.~\eqref{eq:D-flow-general} vanish and the flow equation for $D$ simplifies as
\begin{equation}
\label{eq:D-flow-ordered}
\frac{\dd D}{\dd k} = - \rho_m \frac{\delta^2 }{\delta^2\widetilde{\rho}} \frac{\partial \Gamma}{\partial k}\biggr|_{\substack{\widetilde{\rho}=0\\ \rho=\rho_m}} = - \rho_m \frac{\delta^2 \Delta\Gamma_2 }{\delta^2\widetilde{\rho}} \biggr|_{\substack{\widetilde{\rho}=0\\ \rho=\rho_m}},
\end{equation}
where we used Eqs.~\eqref{eq:wetterich-series} and~\eqref{eq:DeltaGamma-n} with $V(x)$ as in Eq.~\eqref{eq:self-energy-ordered}. In fact, a direct inspection of Eq.~\eqref{eq:dGamma-2} shows that $\Delta \Gamma_2$ contains a term proportional to $\widetilde{\rho}^2$, while any other term $\Delta \Gamma_n$ with $n>2$ generated by the field-dependent function~\eqref{eq:self-energy-ordered} vanishes when evaluated for $\rho =\rho_m$. 
Accordingly, by calculating $\Delta \Gamma_2$ as in Eq.~\eqref{eq:dGamma-2} (see also App.~\ref{app:evaluation-DGamma}), in the long-time limit within which the function $f_D(t)$ (see Eqs.~\eqref{eq:dGamma-2} and~\eqref{eq:fD}) vanishes, and by applying Eq.~\eqref{eq:D-flow-ordered}, we find the equation
\begin{equation}
\frac{\dd D}{\dd k} = -3  k^{d+1} \frac{a_d}{d}\, \frac{KD^2}{Z}\, \left(1 - \frac{\eta_K}{d+2}\right) \frac{\rho_mg^2}{(Kk^2 + m)^4},
\end{equation}
with $m$ given in Eq.~\eqref{eq:effective-mass}. Note that the factor $1 - \eta_K/(d+2)$ comes from the integration over momenta with $\dd R/\dd k$ given by Eq.~\eqref{eq:regulator-derivative}. According to definition~\eqref{eq:anomalous-dimensions}, $\eta_D$ is eventually given by:
\begin{equation}
\eta_D = 3  k^{d+2} \frac{a_d}{d}\, \frac{KD}{Z}\, \left(1 - \frac{\eta_K}{d+2}\right) \frac{\rho_m g^2}{(Kk^2 + m)^4}.
\end{equation}

\subsection{Renormalization of Z}
\label{app:anomalousZ}
Following the general procedure described, e.g., in Refs.~\onlinecite{Canet2007,Sieberer2014}, in order to evaluate the correction to the coefficient $Z$, we express the fields $\phi$ and $\res$ as fluctuations around the homogeneous field $\phi_m$:
\begin{equation}
\label{eq:fluct}
\phi(\rr,t)=\phi_m+\delta\phi(\rr,t), \quad \res(\rr,t)=\delta\res(\rr,t).
\end{equation}
By replacing Eq.~\eqref{eq:fluct} into the field-dependent function $V(x)$ defined in Eq.~\eqref{eq:self-energy-ordered}, we find
\begin{equation}
\label{eq:self-energy-Z}
V(x) = V_1(x) + V_2(x),
\end{equation}
with $V_1$ being linear in the fluctuations $\delta \phi$ and $\delta \res$, i.e., 
\begin{equation}
\label{eq:self-energy-V1}
V_1(x) = -g\phi_m \vartheta(t)
\begin{pmatrix}
\delta\res(x) & \delta\phi(x) \\
\delta\phi(x)	&	 0
\end{pmatrix},  
\end{equation}
while $V_2$ contains only terms quadratic in the fluctuations, i.e., 
\begin{equation}
V_2(x) =  -g\vartheta(t)
\begin{pmatrix}
\delta\phi(x)\delta\res(x) & [\delta\phi(x)]^2/2 \\
[\delta\phi(x)]^2/2	&	 0
\end{pmatrix}.
\end{equation}
When $V$ in Eq.~\eqref{eq:self-energy-Z} is substituted in the expression of $\Delta\Gamma_2$ in Eq.~\eqref{eq:DeltaGamma2-computation-1}, it produces a term which contains a product of two $V_1$ calculated at different spatial and temporal coordinates, generating a quadratic term $\propto \delta\phi\delta\res$, non-local in both spatial and temporal coordinates. 
Since we are interested in the renormalization of $Z$, we can restrict to terms which are local in space and therefore we can use Eq.~\eqref{eq:DeltaGamma2-computation-1} and replace $V$ by $V_1$ in it: this gives (cf. Eq.~\eqref{eq:DeltaGamma2-computation-2})
\begin{widetext}
\begin{align}
\label{eq:DeltaGamma2-V1}
\Delta \Gamma_2 \big|_{V\to V_1}
& = 4 k^{d+1}\frac{a_d}{d}K\left(1-\frac{\eta_K}{d+2}\right)g^2\rho_m\int_{\rr}\int_{t_0}^{+\infty} \dd t\,\dd t' \dd t''\, \bigg\{ 2\, \delta\res(t')\delta\res(t'') \, G_{0C}(k,t',t'')G_{0C}(k,t'',t)G_{0R}(k,t',t) \nonumber \\ 
& \qquad  + \delta\res(t')\delta\phi(t'') \bigg[ G_{0C}(k,t',t'')G_{0R}(k,t,t'')G_{0R}(k,t',t) \nonumber \\
& \qquad + G_{0C}(k,t',t'')G_{0R}(k,t,t'')G_{0R}(k,t',t) + G_{0C}(k,t,'t)G_{0R}(k,t',t'')G_{0R}(k,t'',t) \bigg]\bigg\},
\end{align}
\end{widetext}
where the dependence of the fluctuations on the spatial coordinates $\rr$ has been omitted for simplicity.       
Then, by neglecting the term $\propto \delta\res^2$, which generates only additional irrelevant terms, and by expanding $\delta\phi(t'')$ for $t''\simeq t'$ as $\delta\phi(t'') \simeq \delta \phi(t') + (t''-t')\delta\dot{\phi}(t')$, keeping only the derivative, from Eq.~\eqref{eq:DeltaGamma2-V1} we find 
\begin{align}
\label{eq:DeltaGamma-Z}
& \Delta \Gamma_2 \big|_{V\to V_1} \simeq  4 k^{d+1}\frac{a_d}{d}K\left(1-\frac{\eta_K}{d+2}\right)g^2\rho_m \nonumber \\
& \qquad \qquad \qquad \times \int_{\rr}\int_{t_0}^{+\infty} \dd t' \delta\res(t')\delta\dot{\phi}(t') F_Z(t'), 
\end{align}
where 
\begin{align}
\label{eq:FZ-definition}
F_Z(t') 
& = \int_{t_0}^{+\infty}\dd t\,\dd t'' (t''-t') \nonumber \\
&  \quad\quad\quad \times \bigg[ G_{0C}(k,t',t'')G_{0R}(k,t,t'')G_{0R}(k,t',t) \nonumber \\
&  \quad\quad\quad + G_{0C}(k,t,t'')G_{0R}(k,t',t'')G_{0R}(k,t',t) \nonumber \\
&  \quad\quad\quad + G_{0C}(k,t,t')G_{0R}(k,t',t'')G_{0R}(k,t'',t)\bigg].
\end{align}
The function $F_Z(t)$ can be easily evaluated using Eqs.~\eqref{eq:GK-gaussian} and~\eqref{eq:GR-gaussian}, and  reads
\begin{equation}
\label{eq:FZ-expression}
F_Z(t) = \frac{D}{4\omega_k^4}\left[3 - f_Z(t-t_0)\right],
\end{equation}
where $f_Z(t)$ is a function which vanishes exponentially fast upon increasing $t$ and therefore does not contribute to the renormalization of $Z$ at long times. 
Finally, by replacing Eq.~\eqref{eq:FZ-expression} into Eq.~\eqref{eq:DeltaGamma-Z}, and by comparing the r.h.s.~of Eq.~\eqref{eq:wetterich-series} with its l.h.s.~in which the effective action~\eqref{eq:effective-action} has been inserted, one finds the flow equation for $Z$:
\begin{equation}
\frac{\dd Z}{\dd k}=-3k^{d+1}\frac{a_d}{d}\frac{K D}{Z}\left(1-\frac{\eta_K}{d+2}\right)\frac{g^2\rho_m}{(Kk^2+m)^4}Z,
\end{equation}
where $m$ is given in Eq.~\eqref{eq:effective-mass}. By using the definitions in Eq.~\eqref{eq:anomalous-dimensions}, one thus finds the expression of the anomalous dimension $\eta_Z$:
\begin{equation}
\eta_Z = 3k^{d+2}\frac{a_d}{d}\frac{K D}{Z}\left(1-\frac{\eta_K}{d+2}\right)\frac{g^2\rho_m}{(Kk^2+m)^4}.
\end{equation}

\subsection{Renormalization of K}
\label{app:anomalousK}
The calculation of the flow equation for $K$ proceeds as in the case of $Z$ discussed in the previous section, i.e., we expand the field $\phi$ around the homogeneous configuration as in Eq.~\eqref{eq:fluct}. This renders the same field-dependent function $V(x)$ as in Eq.~\eqref{eq:self-energy-Z}, containing a term $V_1$ linear in the fluctuations which --- when inserted in the expression~\eqref{eq:DeltaGamma2-computation-1} for $\Delta\Gamma_2$ --- generates quadratic terms which are non-local in spatial and temporal coordinates. 
It is convenient to define $K$ as follows~\cite{Canet2007}:
\begin{equation}
K = \mathcal{N} \frac{\partial}{\partial p^2} \frac{\delta^2 \Gamma}{\delta\res(t,-\pp)\delta\phi(t,\pp)}\biggr|_{\substack{\pp=0\\ \delta\res=\delta\phi =0}},
\end{equation}
where $\mathcal{N}$ is a normalization factor formally given by $\mathcal{N} = (2\pi)^d/[\delta^{(d)}(q=0)\delta(t=0)]$, and $\pp$ is a given momentum, eventually vanishing. By taking the total derivative with respect to $k$ of the previous expressions, we find
\begin{align}
\label{eq:K-flow}
\frac{\dd K}{\dd k}
& = \mathcal{N} \frac{\partial}{\partial p^2} \frac{\delta^2 }{\delta\res(t,-\pp)\delta\phi(t,\pp)}\frac{\partial \Gamma}{\partial k}\biggr|_{\substack{\pp=0\\ \delta\res=\delta\phi =0}} \nonumber \\
& =\mathcal{N} \frac{\partial}{\partial p^2} \frac{\delta^2\Delta\Gamma_2\big|_{V\to V_1} }{\delta\res(t,-\pp)\delta\phi(t,\pp)}\biggr|_{\substack{\pp=0\\ \delta\res=\delta\phi =0}},
\end{align}
where in the last equality we used Eq.~\eqref{eq:wetterich-series} and the fact that the sole non-trivial contribution comes from the part of $\Delta\Gamma_2$ (indicated as $\Delta\Gamma_2\big|_{V\to V_1}$ in the previous equation) involving the product of two $V_1$ (see App.~\ref{app:anomalousZ}). From Eq.~\eqref{eq:DeltaGamma2-computation-1}, we find with some simple calculations:
\begin{align}
\label{eq:DeltaGamma2-K}
&\Delta \Gamma_2\big|_{V\to V_1}	= \nonumber \\
& \quad \frac{1}{2} \int_{t,t',t'',\qq,\qq'} \!\!\!\!\! \text{tr}\biggr[ G_0(q,t,t') V(t',\qq-\qq') G_0(q',t',t'') \nonumber \\
& \quad \quad \quad \times V(t'',\qq'-\qq) G_0(q,t'',t)\frac{\dd R}{\dd k}(q)\biggr] \nonumber \\
& \simeq 2g^2\rho_m\int_{t',\qq,\qq'}  \!\!\!\!\!\!\!\! \delta\res(\qq-\qq',t')\delta\phi(\qq'-\qq,t') F_K(q,q',t') \frac{\dd R(q)}{\dd k},
\end{align}
where $V(t,\qq) = \int_\rr \ee^{-i\qq\cdot\rr} V(t,\rr)$ and in the last step we retained only the part of the fields which is local in time by expanding them as $\delta \phi(t'')\simeq \delta\phi(t')$ for $t'' \simeq t'$. In the last equality of Eq.~\eqref{eq:DeltaGamma2-K}, we also discarded the term proportional to $\delta\res^2$, which does not contribute to the renormalization of $K$. The function $F_K(q,q',t')$ in Eq.~\eqref{eq:DeltaGamma2-K} is defined as
\begin{align}
\label{eq:FK-definition}
& F_K(q,q',t') = \nonumber \\
&  \qquad\quad\int_{t_0}^{+\infty}\dd t\,\dd t'' \bigg[ G_{0C}(q',t',t'')G_{0R}(q,t,t'')G_{0R}(q,t',t) \nonumber \\
&  \quad\quad\quad + G_{0C}(q,t,t'')G_{0R}(q',t',t'')G_{0R}(q,t',t) \nonumber \\
&  \quad\quad\quad + G_{0C}(q,t,t')G_{0R}(q',t',t'')G_{0R}(q,t'',t)\bigg].
\end{align}
Then, combining Eqs.~\eqref{eq:K-flow} and~\eqref{eq:DeltaGamma2-K}, one finds
\begin{equation}
\label{eq:DeltaGamma-2-second-variation}
\frac{\delta^2 \Delta\Gamma_2\big|_{V\to V_1}}{\delta\res(t,-\pp)\delta\phi(t,\pp)} = \frac{2g^2\rho_m}{\mathcal{N}} \int_\qq F_K(q,|\qq-\pp|,t)\frac{\dd R (q)}{\dd k}.
\end{equation}
In order to evaluate Eq.~\eqref{eq:K-flow}, we need to retain the contribution proportional to $p^2$ from $F_K(q,|\qq-\pp|,t)$ defined in Eq.~\eqref{eq:FK-definition}. 
To this end, we define the function $P(q)\equiv K[q^2+(k^2-q^2)\vartheta(k^2-q^2))]$ and note that $G_{0R,0C}(q,t,t')$ depend on $q$ via $P(q)$, as their explicit expression is given by Eqs.~\eqref{eq:GK-definition} and~\eqref{eq:GR-definition} with $\omega_q$ replaced by $\omega_{k,q}=P(q) +\tau$ given in Eq.~\eqref{eq:modified-dispersion}. For a generic function of $P(q)$ one can write 
\begin{equation}
\frac{\partial^2}{\partial q_i \partial q_j} = \frac{\partial P(q)}{\partial q_i}\frac{\partial P(q)}{\partial q_j} \frac{\partial^2}{\partial P^2} + \frac{\partial^2 P}{\partial q_i \partial q_j} \frac{\partial }{\partial P},
\end{equation}
where $i,j = 1,\dots, d$ label the components of the momenta $q_i$. A simple calculation shows that
\begin{align}
\frac{\partial P(q)}{\partial q_i} & = 2K q_i\left[1-\vartheta(k^2-q^2) \right], \nonumber \\
\frac{\partial^2 P(q)}{\partial q_i\partial q_j} & = 2K \delta_{ij}\left[1-\vartheta(k^2-q^2)\right] + 4Kq_iq_j\delta(k^2-q^2),
\end{align}
and therefore, all contributions proportional to $1-\vartheta(k^2-q^2)$  vanish when inserted into the integral in the r.h.s.~of Eq.~\eqref{eq:DeltaGamma-2-second-variation}, because they multiply the term $\propto \vartheta(k^2-q^2)$ contained in $\dd R/\dd k$ (see Eq.~\eqref{eq:regulator-derivative}). Accordingly, by discarding these contributions, the derivatives $\partial^2G_{0R,0C}(q,t,t')/\partial q_i \partial q_j$ which are involved in the expansion of the function $F_K$ in the integrand of Eq.~\eqref{eq:DeltaGamma-2-second-variation} can be effectively replaced by
\begin{equation}
\label{eq:greens-second-derivative}
\frac{\partial^2G_{0R/0K}(q,t,t')}{\partial q_i\partial q_j} \mapsto \frac{\partial G_{0R/0K}(q,t,t')}{\partial P(q)} 4Kq_iq_j\delta(k^2-q^2),
\end{equation}
where, from Eqs.~\eqref{eq:GR-gaussian} and~\eqref{eq:GK-gaussian} with $\omega_q \to \omega_{k,q}$ given in Eq.~\eqref{eq:modified-dispersion}, we have
\begin{align}
\frac{\partial G_{0R}(q, t, t')}{\partial P(q)} & = -(t-t')G_{0R}(q, t, t'), \label{eq:GR-derivative}\\
\frac{\partial G_{0C}(q, t, t')}{\partial P(q)} & = -\frac{D}{\omega_{k,q}} \biggr[ \left( \frac{1}{\omega_{k,q}}+|t-t'| \right)\ee^{-\omega_{k,q}|t-t'|} \nonumber\\
& +\left(\frac{1}{\omega_{k,q}} + t+t'\right)\left(\frac{\omega_{k,q}}{D\tau_0}-1\right)\ee^{-\omega_{k,q}(t+t')}\biggr].\label{eq:GK-derivative}
\end{align}
Accordingly, terms proportional to $p^2$ in the Taylor expansion of $F_K(q,|\qq-\pp|,t)$ can be obtained by using Eqs.~\eqref{eq:FK-definition}, \eqref{eq:greens-second-derivative}, \eqref{eq:GR-derivative} and~\eqref{eq:GK-derivative}, and they eventually read
\begin{equation}
\label{eq:FK-Taylor}
\frac{1}{2}\sum_{i,j=1}^d p_ip_j\frac{\partial^2 F_K(q,q,t)}{\partial q_i\partial q_j} = 2K(\qq\cdot\pp)^2\delta(k^2-q^2)\widetilde{F}_K(q,t), 
\end{equation}
with 
\begin{align}
\label{eq:FKtilde-definition}
&\widetilde{F}_K(q,t') = \nonumber \\
&  \quad\quad\int_{t_0}^{+\infty}\dd t\,\dd t'' \bigg[ \frac{\partial G_{0C}(q,t',t'')}{\partial P(q)}G_{0R}(q,t,t'')G_{0R}(q,t',t) \nonumber \\
&  \quad\quad\quad + G_{0C}(q,t,t'')\frac{G_{0R}(q,t',t'')}{\partial P(q)}G_{0R}(q,t',t) \nonumber \\
&  \quad\quad\quad + G_{0C}(q,t,t')\frac{\partial G_{0R}(q,t',t'')}{\partial P(q)}G_{0R}(q,t'',t)\bigg].
\end{align}
Then, by inserting Eq.~\eqref{eq:FK-Taylor} into Eq.~\eqref{eq:DeltaGamma-2-second-variation}, and by using the fact that, from Eq.~\eqref{eq:regulator-derivative}
\begin{equation}
\frac{\dd R_k}{\dd k}\delta(k^2-q^2)=\frac{1}{2}K\delta(k-q),
\end{equation}
as well as the identity for the $d$-dimensional integral of a rotational-invariant function $f(q)$
\begin{equation}
\int \dd^dq\, (\qq\cdot \pp)^2 f(q)=\frac{p^2}{d}\int \dd^dq\, q^2\,f(q),
\end{equation}
we find
\begin{equation}
\label{eq:DeltaGamma2-FK}
\frac{\partial^2}{\partial p^2}\frac{\delta^2 \Delta\Gamma_2\big|_{V_1^2}}{\delta\res(t,-\pp)\delta\phi(t,\pp)} \biggr|_{\pp=0} \!\!\!\!\!
= 2\frac{a_d}{d}k^{d+1}\frac{g^2\rho_m}{\mathcal{N}} K^2\widetilde{F}_K(k,t).
\end{equation}
Finally, a lengthy but straightforward evaluation of $\widetilde{F}_K(k,t)$, using Eqs.~\eqref{eq:FK-definition}, \eqref{eq:GR-gaussian} and~\eqref{eq:GK-gaussian} with $\omega_q \to \omega_{k,q}$ (see Eq.~\eqref{eq:modified-dispersion}), yields
\begin{equation}
\label{eq:FKtilde-expression}
\widetilde{F}_K(k,t) = -\frac{D}{Z\omega_k^4}\left[1-f_K(t)\right],
\end{equation}
where $\omega_{k} = \omega_{q=k}$ and $f_K(t)$ is a function which vanishes exponentially fast upon increasing $t$ and therefore does not contribute to the renormalization of $K$ at long times. By inserting Eqs.~\eqref{eq:FKtilde-expression} and~\eqref{eq:DeltaGamma2-FK} into Eq.~\eqref{eq:K-flow}, we finally find the flow equation for $K$, which reads 
\begin{equation}
\frac{\dd K}{\dd k} = -2 k^{d+1}\frac{a_d}{d}\frac{DK^2}{Z}\frac{g^2\rho_m}{(Kk^2+m)^4},
\end{equation}
and, according to Eq.~\eqref{eq:anomalous-dimensions}, the anomalous dimension $\eta_K$ reads
\begin{equation}
\eta_K = 2 k^{d+2}\frac{a_d}{d}\frac{DK}{Z}\frac{g^2\rho_m}{(Kk^2+m)^4}.
\end{equation}

\section{Flow equations}
\label{app:FRGbetafunct}
In this Appendix we report the explicit form of the flow equations derived from the effective action $\Gamma$ in Eq.~\eqref{eq:effective-action} with the potential $\mathcal{U}$ in Eq.~\eqref{eq:potential-ordered} and $\lambda \neq 0$.
These equations can be derived by repeating the calculations presented in Apps.~\ref{app:beta_ordered} and~\ref{app:anomalous-dimensions} but by keeping $\lambda$ finite; here we report only the final result of this somewhat lengthy calculation.
The flow equations for the couplings $\widetilde{m}$, $\widetilde{g}$ and $\widetilde{\lambda}$, defined in Eq.~\eqref{eq:couplings-dimensionless}, turn out to be
\begin{widetext}
\begin{align}
k\frac{\dd \meff }{ \dd k}  & = (-2 + \eta_K)\meff + \left( 1 - \frac{\eta_K}{d+2} \right) \frac{2\geff}{(1+\meff)^2} \left[ 1 + \frac{3}{2} \left(\frac{\meff\leff}{\geff^2}\right)^2 +\frac{3\meff}{1+\meff} \left( 1 + \frac{\meff\leff}{\geff^2} \right)^2 \right], \label{eq:m-RG}\\
k\frac{\dd \geff}{ \dd k} & = g\left[d-4 +2\eta_K + \left( 1 - \frac{\eta_K}{d+2} \right)\frac{6 g}{(1+\meff)^3}\left( 1 + \frac{\meff\leff}{\geff^2} \right)^2\right] + \left( 1 - \frac{\eta_K}{d+2} \right) \frac{\leff}{(1+\meff)^2}\left(-2 + 3 \frac{\meff\leff}{\geff^2}\right), \label{eq:g-RG}\\
k\frac{\dd \leff}{\dd k} & = \leff\left[ 2d - 6 + 3\eta_K + 30 \left( 1 - \frac{\eta_K}{d+2} \right) \frac{\geff}{(1+\meff)^3}\left( 1 + \frac{\meff\leff}{\geff^2} \right)\right] - 18\left( 1 - \frac{\eta_K}{d+2} \right) \frac{\geff^2}{(1+\meff)^4}\left( 1 + \frac{\meff\leff}{\geff^2} \right), \label{eq:lambda-RG}
\end{align}
\end{widetext}
while the anomalous dimensions $\eta_K, \eta_D, \eta_Z$ and $\eta_0$, defined, respectively, in Eqs.~\eqref{eq:anomalous-dimensions} and~\eqref{eq:Z0-eq-dimensionless}, read
\begin{widetext}
\begin{align}
\eta_K & = \frac{3 \meff \geff}{(1+\meff)^4}\left( 1 + \frac{\meff\leff}{\geff^2} \right)^2, \label{eq:etaK-RG}\\
\eta_Z & = \eta_D = \left( 1 - \frac{\eta_K}{d+2} \right)\frac{9 \meff \geff}{2(1+\meff)^4}\left( 1 + \frac{\meff\leff}{\geff^2} \right)^2, \label{eq:etaZ-RG}\\
\eta_0 & = - \left( 1 - \frac{\eta_K}{d+2} \right)\frac{\geff}{(1+\meff)^3}\left[1  + \frac{3}{2}\left(\frac{\meff\leff}{\geff^2}\right)^2 + \frac{9 \meff}{2(1+\meff)}\left( 1 + \frac{\meff\leff}{\geff^2} \right)^2\right]. \label{eq:eta0-RG}
\end{align}
\end{widetext}
Setting to zero Eqs.~\eqref{eq:m-RG}, \eqref{eq:g-RG}, \eqref{eq:lambda-RG}, we find numerically (using Wolfram Mathematica) the following fixed point values of the rescaled couplings (up to the second  significative digit): 
\begin{equation}\label{eq:FPvaluess}
\tilde{m}^*\simeq0.30, \quad\tilde{g}^*\simeq0.26, \quad\tilde{\lambda}^*\simeq0.04.
\end{equation}
The linearization of the flow equations \eqref{eq:m-RG}, \eqref{eq:g-RG}, and \eqref{eq:lambda-RG} around the fixed point values  given in Eq.~\eqref{eq:FPvaluess}, determines the associated stability matrix, and from the inverse of its negative eigenvalue (see for instance Ref. \onlinecite{Amit/Martin-Mayor}), we find the critical exponent $\nu$ reported in Sec.~V.~C.
The values $\eta^*_{K,Z,0}$ of the anomalous dimensions  at the fixed point are  found  by replacing directly Eq.~\eqref{eq:FPvaluess} into the expressions \eqref{eq:etaK-RG}, \eqref{eq:etaZ-RG}, and \eqref{eq:eta0-RG}.


\bibliography{biblio}

\end{document}